\documentclass[12pt]{article}
\usepackage{amssymb}
\usepackage{epsfig}
\usepackage{amsthm}
\usepackage{amsmath}
\usepackage{amsfonts}
\usepackage{color}
\newcommand{\field}[1]{\mathbb{#1}}
\newcommand{\R}{\field{R}}

\newcommand{\N}{\field{N}}
\newcommand{\Z}{\field{Z}}

\newcommand{\E}{\field{E}}
\def\qed{\hfill$\diamondsuit$}

\theoremstyle{example}
\theoremstyle{remark}
\theoremstyle{lemma}
\theoremstyle{definition}
\theoremstyle{corol}
\theoremstyle{proposition}
\theoremstyle{condition}
\theoremstyle{assumption}
\newtheorem{assumption}{\n{Assumption}}[section]
\newtheorem{theorem}{\n{Theorem}}[section]

\newtheorem{remark}{\n{Remark}}[section]

\newtheorem{proposition}{\n{Proposition}}[section]

\font\n=cmcsc10
\def\cov{{\mbox{cov}}}
\def\var{{\mbox{var}}}
\def\cum{{\mbox{cum}}}
\begin{document}

\newpage

\begin{center}
 {\bf\Large A self-normalized approach to confidence interval construction  in time  series  \footnote{Xiaofeng Shao is Assistant Professor, Department of Statistics, University of Illinois, at Urbana-Champaign,  Champaign, IL 61820 (e-mail: xshao@uiuc.edu). I am grateful to two referees for their constructive comments that led to substantial improvement of the paper.  I would like to thank Xuming He, Roger Koenker and Xianyang Zhang for helpful suggestions on an earlier version.  The research is  supported in
 part by NSF grants DMS-0804937 and DMS-0724752. All errors are attributed solely to the author. }}
\end{center}
\centerline{\today}
  \centerline{\textsc{Xiaofeng Shao}}

\date{}

\centerline {\it  University of Illinois at
Urbana-Champaign}

{\bf Abstract}: We propose a new method to construct confidence intervals for quantities that are associated with a stationary time series, which  avoids direct estimation of the asymptotic variances. Unlike the existing tuning-parameter-dependent approaches, our method has the attractive convenience of being free of choosing any user-chosen number or smoothing parameter.  The interval is constructed on the basis of an asymptotically distribution-free self-normalized statistic,  in which the normalizing matrix is computed using recursive estimates. Under mild conditions, we establish the theoretical validity of our method for a broad class of statistics that are functionals of the empirical distribution of fixed or growing dimension.  From a practical point of view, our method is conceptually simple, easy to implement and can be readily used by the practitioner.
 Monte-Carlo simulations are conducted to compare the finite sample performance of the new method with those delivered by the normal approximation and the block bootstrap approach.

 \bigskip


\newpage

\section{Introduction}

In time series analysis, constructing a confidence interval for an unknown quantity is often difficult owing to dependence.
 For example, for a stationary time series $(X_t)_{t\in\Z}$, suppose the quantity of interest, $\theta$, is the median of the marginal distribution of $X_1$ (which is denoted as $med(X_1)$). Once we observe
the data $(X_1,\cdots,X_n)$, a natural estimator of $\theta$ is  $\hat{\theta}_n=med(X_1,\cdots,X_n)$. Under suitable weak dependence conditions, one can show that
\begin{eqnarray}
\label{eq:an}
\sqrt{n}(\hat{\theta}_n-\theta)\rightarrow_{D} N(0,\sigma^2),
\end{eqnarray}
where ``$\rightarrow_{D}$" stands for convergence in distribution,
\[\sigma^2=\{4 g^2(\theta)\}^{-1}\sum_{k=-\infty}^{\infty}\cov\{1-2{\bf 1}(X_0\le \theta),1-2{\bf 1}(X_k\le \theta)\}\]
 with $g(\cdot)$ being the density function of $X_1$; see B\"uhlmann (2002).
To establish a confidence interval for $\theta$ based on expression (\ref{eq:an}), we need to find a consistent  estimate of $\sigma^2$, which boils down to estimating $g(\theta)$ and the long-run variance of the transformed series $1-2{\bf 1}(X_t\le \theta)$. The consistent estimates of these two quantities both involve a smoothing parameter, the choice of which has been extensively studied in the literature. However, no empirical investigation seems to be done along this line, in part because there are more appealing alternatives, such as
 the moving  block bootstrap method (K\"unsch 1989; Liu and Singh 1992) and the subsampling approach (Politis, Romano and Wolf 1999). The resampling techniques are powerful in that they bypass direct estimation, and the resulting confidence intervals have asymptotically correct coverage probability under appropriate conditions. However, a practical drawback that is associated with these methods is that they all require the selection of a user-chosen parameter, such as the block length in the moving block bootstrap, and the window width in the subsampling method. The empirical coverage probability can be sensitive to the choice of these user-chosen numbers. Although the methods that address the optimal choice of the tuning parameters are available (Hall, Horowitz and Jing 1995;  Politis et al. 1999), they are usually rather {\it ad hoc}, or involve another user-chosen number and require very expensive
 computation. The confidence interval can also be constructed by using the blockwise empirical likelihood method (Kitamura 1997), but again one must deal with  the issue of block size selection.

In this paper, we propose a new approach to constructing confidence intervals (regions) for a large class of quantities that are encountered in time series analysis.  The method does not involve  any user-chosen numbers and yields  a confidence interval that has asymptotically correct coverage.
   The interval is constructed on the basis of a self-normalized statistic, where the normalization matrix is formed by using recursive estimates.
   The self-normalized method proposed is an extension of Lobato (2001) from the sample
    autocovariances to more general approximately linear statistics. It also relates to recent work on fixed-b asymptotics in the econometrics literature (Kiefer, Vogelsang and Bunzel 2000; Kiefer and Vogelsang 2002b,2005 among others). As an important methodological contribution,  the new approach can be used to construct confidence intervals and to test hypotheses based on the approximately linear statistic that has a non-differentiable influence function (e.g. the sample median), to which all the aforementioned works are not directly applicable. Further, it can be extended to confidence interval construction for the quantity that is a functional of the joint  distribution of $\{X_t\}_{t\in\Z}$, which is of interest in  spectral analysis.

We now introduce some notation. For a column vector $x = (x_1,
\cdots, x_q)'\in \R^q$, let $|x| = (\sum_{j=1}^q x_j^2)^{1/2}$.
Let $\xi$ be a random vector. Write $\xi \in {\cal L}^p$ ($p > 0$)
if $\|\xi\|_p := [\E(|\xi|^p )]^{1/p} < \infty$ and let $\|\cdot\|
= \|\cdot\|_2$. The symbols $O_p(1)$ and $o_p(1)$ signify being bounded
in probability and convergence to zero in probability, respectively. Denote by $\lfloor a\rfloor$ the integer part of $a$.

 The paper is organized as follows. Section~\ref{sec:method}
introduces the main idea of confidence interval construction for quantities that are functionals of the finite dimensional marginal distribution,
describes its connection to the fixed-b approach and proposes a new test for non-correlation. Section~\ref{sec:extension} extends the applicability of our method to quantities that are functionals of the whole joint distribution of the series. In Section~\ref{sec:sim},  simulation results are presented to examine the finite sample performance
of the new method in comparison with the standard and block bootstrap approaches. Section~\ref{sec:con} concludes and technical details are relegated to the appendix.

\section{Methodology}
\label{sec:method}

In this section, we confine our discussion to quantities that can be expressed as  functionals of the $m$-dimensional marginal distribution of $(X_t)_{t\in\Z}$, where  $m$ is a fixed but arbitrary integer. In other words, let $\theta=T(F_m)$, where
 $F_m$ is the marginal distribution of $Y_t=(X_t,\cdots,X_{t+m-1})'$ and $T$ is a functional that takes values in $\R^q$. Let $N=n-m+1$ and
 $\rho_m^{N}=N^{-1}\sum_{t=1}^N \delta_{Y_t}$ be the empirical distribution, where $\delta_y$ denotes the point mass at $y\in \R^m$. A natural estimator of $\theta$ is $\hat{\theta}_N=T(\rho_m^N)$. We shall focus on the class of  statistics that are approximately linear in this section. For an approximately linear statistic $T(\rho_m^N)$, it admits the following expansion in a neighborhood of $F_m$, i.e.
\begin{eqnarray}
\label{eq:expansion}
T(\rho_m^N)=T(F_m)+N^{-1}\sum_{t=1}^{N} IF(Y_t;F_m)+R_N,
\end{eqnarray}
 where $IF(Y_t;F_m)$ is the influence function of $T$ (Hampel, Ronchetti, Rousseeuw and Stahel 1986) defined by
 \[IF(y;F_m)=\lim_{\epsilon\downarrow 0} \left[\frac{T\{(1-\epsilon)F_m+\epsilon\delta_y\}-T(F_m)}{\epsilon}\right]\]
 and $R_N$ is the remainder term. For example, the $(m-1)$-th ($m\in\N$) lag autocovariance and autocorrelation, which are denoted by $\gamma(m-1)=\cov(X_0,X_{m-1})$ and $\rho(m-1)=\gamma(m-1)/\gamma(0)$  respectively, depend only on $F_m$. Their sample estimates  $$\tilde{\gamma}_n(m-1)=(n-|m-1|)^{-1}\sum_{t=1}^{n-|m-1|}(X_t-\bar{X}_n)(X_{t+|m-1|}-\bar{X}_n),$$
  where  $\bar{X}_n=n^{-1}\sum_{t=1}^{n}X_t$,
 and $\tilde{\rho}_n(m-1)=\tilde{\gamma}_n(m-1)/\tilde{\gamma}_n(0)$  are functionals of $\rho_m^N$. Note that the commonly used estimates $\hat{\gamma}_n(m-1)=n^{-1}\sum_{t=1}^{n-|m-1|}(X_t-\bar{X}_n)(X_{t+|m-1|}-\bar{X}_n)$ and $\hat{\rho}_n(m-1)=\hat{\gamma}_n(m-1)/\hat{\gamma}_n(0)$ differ from $\tilde{\gamma}_n(m-1)$ and $\tilde{\rho}_n(m-1)$  by a constant factor, and it is easy to see that these two definitions are asymptotically equivalent for a fixed $m$.  See K\"unsch (1989) for more examples of approximately linear statistics, such as various location and scale estimators for the marginal distribution of $X_1$, von Mises statistics and M-estimators of time series models.

  Under expansion (\ref{eq:expansion}) and some regularity conditions that ensure the negligibility of $R_N$,
 $\sqrt{N}(\hat{\theta}_N-\theta)\rightarrow_{D} N\{0,\Sigma(F_m)\}$, where 
 $$\Sigma(F_m)=\sum_{k=-\infty}^{\infty}\cov\{IF(Y_0;F_m),IF(Y_k;F_m)\}$$ 
 is the long-run variance matrix of the stationary process $\{IF(Y_t;F_m)\}_{t\in\Z}$. Equivalently, $\Sigma(F_m)$ is the spectral density matrix of $\{IF(Y_t;F_m)\}$ evaluated at zero frequency (up to a constant factor).
As shown in the example of the median,
$\Sigma(F_m)$ could contain some nuisance parameters, which render consistent estimation of $\Sigma(F_m)$ a difficult task.

 To motivate our proposal, we consider $\hat{\theta}_{\lfloor rN\rfloor}=T(\rho_m^{\lfloor rN\rfloor})$ for $r\in (0,1]$, which estimates $\theta$  on
 the basis of the subsample of first $\lfloor rN\rfloor $ observations of $Y_t$, i.e. $(Y_1,\cdots,Y_{\lfloor rN\rfloor})$.
 Analogously to equation (\ref{eq:expansion}), we have
\[T(\rho_m^{\lfloor rN\rfloor})=T(F_m)+(\lfloor rN\rfloor)^{-1}\sum_{t=1}^{\lfloor rN\rfloor}IF(Y_t;F_m)+R_{\lfloor rN\rfloor}.\]
Let ${\cal D}[0,1]$ be the space of functions on $[0,1]$ which are right continuous and have left limits, endowed
with the Skorokhod topology (Billingsley 1968). Denote by ``$\Rightarrow$" weak convergence in ${\cal D}[0,1]$.
Our method hinges on the following two assumptions.
\begin{assumption}{\rm
\label{as:as1}
Assume that $\E\{IF(Y_t;F_m)\}=0$ and
\begin{eqnarray}
\label{eq:ip}
N^{-1/2}\sum_{t=1}^{\lfloor rN\rfloor} IF(Y_t;F_m)\Rightarrow \Delta B_q(r),
\end{eqnarray}
where $\Delta$ is a $q\times q$ lower triangular matrix with nonnegative diagonal entries and $B_q(\cdot)$ is a $q$-dimensional vector of independent Brownian motions. Assume that $\Delta\Delta'=\Sigma(F_m)$ is positive definite. }

\end{assumption}

\begin{assumption}
\label{as:as2}
{\rm  Assume that $R_N=o_p(N^{-1/2})$ and $N^{-2}\sum_{t=1}^{N}|tR_t|^2=o_p(1)$.
 %\lfloor rN\rfloor R_{\lfloor rN\rfloor}=o_p(N^{1/2})$ uniformly in $r\in (0,1]$.
}
\end{assumption}

Let $W_N=N^{-2}\sum_{t=1}^{N}t^2(\hat{\theta}_t-\hat{\theta}_N)(\hat{\theta}_t-\hat{\theta}_N)'$. Denote 
\[V_q=\int_{0}^1 \{B_q(r)-rB_q(1)\}\{B_q(r)-rB_q(1)\}' dr~\mbox{and}~U_q=B_q(1)'V_q^{-1}B_q(1).\]
The upper critical values of $U_q$ have been tabulated by Lobato (2001) for $q=1,\cdots,20$.

\begin{theorem}
\label{th:main1}
Under Assumptions \ref{as:as1}  and \ref{as:as2}, we have that
\begin{eqnarray}
\label{eq:main}
N(\hat{\theta}_N-\theta)' W_N^{-1} (\hat{\theta}_N-\theta) \rightarrow_{D} U_q.
\end{eqnarray}
So a $100(1-\alpha)\%$ confidence region for $\theta$ is
\[\{\theta: N(\hat{\theta}_N-\theta)' W_N^{-1} (\hat{\theta}_N-\theta)\le U_{q,\alpha}\},\]
where $U_{q,\alpha}$ is the $100(1-\alpha)$th percentile of the distribution for $U_q$.

\end{theorem}

\noindent Proof of Theorem~\ref{th:main1}: Under Assumption~\ref{as:as1} and $R_N=o_p(N^{-1/2})$, we have $N^{1/2}(\hat{\theta}_N-\theta)\rightarrow_{D} \Delta B_q(1)$. Let $\mbox{S}_{IF}(t)=\sum_{j=1}^{t}IF(Y_j;F_m)-(t/N)\sum_{j=1}^{N}IF(Y_j;F_m)$. Then we can write
\begin{eqnarray*}
t(\hat{\theta}_t-\hat{\theta}_N)=t(\hat{\theta}_t-\theta)-\frac{t}{N}N(\hat{\theta}_N-\theta)=\mbox{S}_{IF}(t)+\left(tR_t-\frac{t}{N}N R_N\right), ~t=1,\cdots,N,
\end{eqnarray*}
which implies that $W_N=N^{-2}\sum_{t=1}^{N} \mbox{S}_{IF}(t) \mbox{S}_{IF}(t)' + o_p(1)$ under Assumption~\ref{as:as2}.
It then follows from Assumption~\ref{as:as1} and the continuous mapping theorem that $W_N\rightarrow_{D} \Delta V_q \Delta'$, which is joint with
$N^{1/2}(\hat{\theta}_N-\theta)\rightarrow_{D} \Delta B_q(1)$. Since $\Delta$ is invertible, the stated result is obtained.
\qed

\begin{remark}{\rm
Assumption~\ref{as:as1} is not primitive. To give primitive conditions, one can  resort to mixing assumptions; see Phillips (1987), Assumption 2.1, which is originally due to Herrndorf (1984). Specifically, condition (\ref{eq:ip}) holds if
\[\E|IF(Y_t;F_m)|^{\beta}<\infty,~\sum_{k=1}^{\infty}\alpha_k^{1-2/\beta}<\infty~\mbox{for some}~\beta>2, \]
 where $(\alpha_k)_{k\in\N}$ stands for the strong mixing coefficients of $(X_t)$. The mixing assumption is mild and it allows for a wide
 variety of time series models, such as finite order auto-regressive moving average models (Pham and Tran 1985), bilinear models (Pham 1986) and
   generalized auto-regressive conditional heteroscedasticity (GARCH) models (Carrasco and Chen 2002). Or one can impose the near-epoch dependence assumption, which allows processes that are not  mixing; see Davidson (2002) for more details. Assumptions~\ref{as:as2} is also mild. Here we show that it is verifiable for the class of smooth function models. This class is sufficiently wide to include many statistics of practical interest, such as auto-covariance, auto-correlation, the Yule-Walker estimator and other interesting statistics in time series. Let $\mu_Z=\E(Z_t)$, where $Z_t$ is a multivariate stationary time series in $\R^p$, i.e.  $Z_t=(Z_t^{(1)},\cdots,Z_t^{(p)})'$. Let $\bar{Z}_N=N^{-1}\sum_{j=1}^{N}Z_j$ and   $\theta=H(\mu_Z)\in \R$. Then $\hat{\theta}_N=H(\bar{Z}_N)$
  and $IF(Z_t;F_m)=(Z_t-\mu_Z)' \partial H(\mu_Z)/\partial z$. By the mean-value theorem,
 $R_N=(\bar{Z}_N-\mu_Z)'\{\partial H^2(\tilde{Z}_N)/\partial^2 z\} (\bar{Z}_N-\mu_Z)$, where $\tilde{Z}_N=\beta \bar{Z}_N+(1-\beta)\mu_Z$ for some $\beta\in [0,1]$.    Assumption~\ref{as:as2} holds provided that $\|\partial^2 H(z)/\partial^2 z\|_2$ is bounded, $\bar{Z}_N-\mu_Z=o_p(N^{-1/4})$ and $N^{-2}\sum_{t=1}^{N} t^2\E |(\bar{Z}_{t}-\mu_Z)|^4=o(1)$, the latter two of which hold if $Z_t\in {\cal L}^4$ and
 $\sum_{k_1,\cdots,k_j\in\Z}|\cum(Z_{0}^{(p_0)},Z_{k_1}^{(p_1)},\cdots,Z_{k_j}^{(p_{j})})|<\infty$ for  any $(p_0,\cdots,p_j)\in \{1,\cdots,p\}^{j+1}$, $j=1,2,3$. Here for a $p\times p$ matrix, $\|A\|_2$ denotes the matrix norm induced by the vector norm $\|z\|_2=(\sum_{j=1}^{p}z_j^2)^{1/2}$.
}
\end{remark}

 By introducing a random normalization matrix $W_N$, which is proportional to $\Sigma(F_m)$, our proposal avoids the thorny issue of estimating $\Sigma(F_m)$ explicitly
and our  statistic is asymptotically pivotal. The idea of using random normalization is not new, and it has been applied by Lobato (2001) to the problem of a non-correlation test. However, the formulation in Lobato (2001) is
  tailored to the testing problem, whereas our method is developed under a more general framework. A distinctive feature of our method is that we use recursive estimates of the quantity of interest in the formation of the normalization matrix. In particular, the normalization matrix in Lobato (2001) is an explicit function of the  cumulative sum  (CUSUM) process corresponding to the sample autocovariances, whereas ours  involves  recursive estimates since the CUSUM process for the influence function is typically unobserved and may involve unknown nuisance parameters.
   From the proof of Theorem~\ref{th:main1}, we see that the use of recursive estimates allows us to express the CUSUM  process on the basis of the influence function as the difference between the process $\{t(\hat{\theta}_t-\hat{\theta}_N)\}_{t=1}^{N}$ and a negligible remainder term.
  The use of recursive estimates  in normalization (standardization) has also been considered by Kuan and Lee (2006) and Lee (2007). However,  their discussions were restricted to the robust testing context, where the statistic of interest is a function of residuals and the use of recursive residuals can remove the estimation effect. As pointed out by a referee, the use of recursive estimates that are computed on a sequence of increasing  subsamples of the series is related to the notion of ``scanning" that was proposed in McElroy and Politis (2007), in which a scan was defined as a collection of $n$ block subsamples of the sequence of $\{X_1,\cdots,X_n\}$, which contains $n$ nested blocks, each of which has size $k$, for $k=1,\cdots,n$.  The one considered in our paper basically corresponds to a forward scan, i.e., $\{(X_1),(X_1,X_2),(X_1,X_2,X_3),\cdots,(X_1,\cdots,X_n)\}$. A natural question is whether the proposed method would work for other scans. It seems that other scans may work with suitable modification of the distribution theory but the practical gain is not clear. We leave this for future investigation.

\subsection{Self-normalization versus fixed-b approach}

 In theory, we can replace $W_N$ with any smooth functional of the process $\{\lfloor rN\rfloor (\hat{\theta}_{\lfloor rN\rfloor}-\hat{\theta}_{N}), r\in (0,1]\}$.  The asymptotic distribution of the resulting statistic is pivotal and its percentiles can be obtained from simulations. Our particular choice of $W_N$ is somewhat arbitrary and is in part influenced by
     Lobato's (2001) proposal, which  is closely linked to  the fixed-b asymptotic scheme that was considered by Kiefer, Vogelsang and their co-authors (Kiefer et al. 2000; Kiefer and Vogelsang 2002a, 2002b, 2005; Bunzel, Kiefer and Vogelsang 2001). To elucidate their connections, we focus on the simple case $\theta=\E(X_1)$. In this case, $\hat{\theta}_t=t^{-1}\sum_{j=1}^t X_j$, $W_n=n^{-2}\sum_{t=1}^{n} \left\{\sum_{j=1}^{t}(X_j-\bar{X}_n)\right\}^2$ and expansion (\ref{eq:expansion}) holds with a vanishing remainder term.
 Equation (\ref{eq:main}) reduces to $n(\bar{X}_n-\theta)^2/W_n \rightarrow_{D} U_1$, which has been discussed in section 2 of Lobato (2001). To construct a confidence interval (or to perform hypothesis testing) for $\E(X_1)$, a standard approach is to find a consistent estimate for  $\lim_{n\rightarrow\infty} \{n\var(\bar{X}_n)\}=\sum_{j=-\infty}^{\infty}\gamma(j)$. The commonly used lag window estimate  admits the form \[\sum_{j=-(n-1)}^{n-1}K\{j/(bn)\} \hat{\gamma}_n(j),\]
 where $K(\cdot)$ is the kernel function and $bn$ is the bandwidth. In the standard asymptotic regime, the ratio of the bandwidth to the sample size $b\rightarrow 0$ as $n\rightarrow\infty$ and the inference is based on the limiting normal or $\chi^2$ distribution, whereas $b\in (0,1]$ is held constant in the fixed-b asymptotics and the limiting distribution is nonstandard depending on the kernel function and $b$. Kiefer and Vogelsang (2002a) showed that $2 W_n$ is equal to  the lag window estimate when $K(\cdot)$ is the Bartlett kernel, i.e., $K(x)=(1-|x|){\bf 1}(|x|\le 1)$ and $b=1$.  Therefore, the self-normalized method proposed can be regarded as a special case of the fixed-b approach. It is also worth noting that
 the self-normalized approach differs from that used in the independent and identically distributed (iid) setting, where one typically normalizes with
 the sample variance (Lai, de la Pena and Shao 2009). For a stationary time series, the sample variance is no longer suitable as the normalization factor, since the long run variance of $X_t$ (i.e., $\sum_{j=-\infty}^{\infty}\gamma(j)$) is the nuisance parameter instead of the marginal variance of $X_t$ (i.e.,  $\gamma(0)$).

 Compared with the standard approach where the normalization (Studentization) factor is a consistent estimator of the asymptotic variance, the self-normalized approach adopts an inconsistent estimator as the normalization factor, which  in a sense corresponds to  `inefficient Studentizing'. In what follows, we describe important implications of the inefficient Studentizing in terms of the size and power behaviors for hypothesis testing and the coverage accuracy for confidence interval construction. For the Gaussian location model, Jansson (2004) showed that
 \begin{eqnarray}
 \label{eq:erp}
 \sup_{x\in\R}\left|P\left(\frac{n(\bar{X}_n-\theta)^2}{W_n}\le x\right)-P\left(U_1\le x\right)\right|=O\{n^{-1}\log (n)\},
 \end{eqnarray}
which was further refined by Sun, Phillips and Jin (2008) under the fixed-b asymptotic framework by dropping the $\log(n)$ term. In the testing context, the implication of equation (\ref{eq:erp})
is that the self-normalized approach controls  the size better than the standard approach, where the corresponding error rejection rate (ERP) is no better than $O(n^{-1/2})$ (Velasco and Robinson 2001). For heuristic and theoretical explanations of the better size property of the self-normalized approach as compared with the standard approach, we refer the reader to  Bunzel et al. (2001) and Sun et al. (2008). When testing for $\gamma(1)=0$,   Lobato (2001) showed that  the local asymptotic power of the self-normalized approach is dominated by the standard approach. Also see  Kiefer et al. (2000) for a similar finding in the context of robust testing for linear regression models with auto-correlated errors. The phenomenon of ``better size but less power" corresponding to the self-normalized approach is also consistent with earlier  Monte Carlo results in Kiefer et al. (2000), Kiefer and Vogelsang (2002b, 2005), Bunzel et al. (2001) and Lobato (2001). See  Sun et al. (2008) for an interesting theoretical explanation of the phenomenon of ``better size but less power" for the fixed-b approach (with the self-normalized approach as a special case) using the loss function argument.

For confidence interval construction, the coverage probability corresponds to the size, so equation (\ref{eq:erp}) implies that
 the coverage accuracy for the self-normalized approach is better than that offered by the standard approach, which is also confirmed in our simulation studies; see Section~\ref{subsec:4}.  The asymptotic analysis and simulation results in  Kiefer and Vogelsang (2005) under the fixed-b framework suggest that for a given kernel function, $b=1$ corresponds to the least size distortion. This supports our choice $b=1$ in confidence interval construction, since our self-normalized statistic is basically a special case of the fixed-b formulation with the Bartlett kernel and $b=1$.  It is possible and in fact quite straightforward  to extend the fixed-b approach to the framework that is described in this section.   We do not pursue  this generalization as there is  no additional methodological and technical difficulty in view of the argument that was used in Kiefer and Vogelsang (2002b, 2005).
  We also note the work by Phillips, Sun and Jin (2004,2006,2007), who estimated the spectral density (or long-run variance) by exponentiating kernels with bandwidth equal to the sample size.  They developed the so-called fixed-exponent asymptotics, which are similar in spirit to the fixed-b asymptotics.

\subsection{A new test for non-correlation}

 Owing to the duality of confidence interval construction and hypothesis testing, we can extend our method to the hypothesis testing context. For example, suppose we are interested in testing
\[H_0:\gamma(1)=\cdots=\gamma(m-1)=0~\mbox{versus}~H_a:\gamma(j)\not=0,~\mbox{for some}~j=1,\cdots,m-1.\]
For $t=1,\cdots,N$, let $\bar{X}_t=(t+m-1)^{-1}\sum_{k=1}^{t+m-1}X_k$ and $\hat{\gamma}_t(j)=(t+m-1)^{-1}\sum_{k=1}^{t+m-1-|j|}(X_k-\bar{X}_{t})(X_{k+|j|}-\bar{X}_t)$ be the estimates of $\E(X_t)$ and $\gamma(j)$ based on the subsample of first $t$ observations of $\{Y_h\}_{h=1}^{N}$. Denote by $c_t(m-1)=\{\hat{\gamma}_t(1),\cdots,\hat{\gamma}_t(m-1)\}'$,
$\tilde{S}_t=t\{c_t(m-1)-c_N(m-1)\}$ for $t=1,\cdots,N$, and $\tilde{J}_{m-1}=N^{-2}\sum_{t=1}^{N}\tilde{S}_t\tilde{S}_t'$.
 Our test statistic is formed as
\[\tilde{T}_{m-1}=N c_N(m-1)' \tilde{J}_{m-1}^{-1} c_N(m-1).\]
Rejection of hypothesis $H_0$ occurs when $\tilde{T}_{m-1}$ is too large, by reference to  upper critical values of $U_{m-1}$. Further let $Z_{kt}=(X_t-\bar{X}_n)(X_{t+k}-\bar{X}_n)$ for $k=1,\cdots,m-1$,
${Z}_t=(Z_{1t},\cdots,Z_{(m-1)t})'$,  $S_t=\sum_{j=1}^{t}\{Z_j-c_N(m-1)\}$ and $J_{m-1}=N^{-2}\sum_{t=1}^{N}S_tS_t'$. Then Lobato's test statistic is
\[T_{m-1}=N c_N(m-1)' J_{m-1}^{-1} c_N(m-1),\]
which has the same distributional limit as $\tilde{T}_{m-1}$.  The difference between  $T_{m-1}$ and $\tilde{T}_{m-1}$ lies in different forms of their normalization matrices; for example  the recursive mean estimate $\bar{X}_t$ is used in $\tilde{J}_{m-1}$, whereas the sample mean $\bar{X}_n$ is used in
$J_{m-1}$. The leads to the difference in their finite sample size and power performance, which will be  elaborated in Section~\ref{subsec:1}.

\section{Theoretical extensions}

\label{sec:extension}

To broaden the applicability of our methodology, we shall consider constructing confidence intervals for quantities  that are functionals of $F_{\infty}$, i.e. the joint distribution of $(X_t)_{t\in\Z}$. To illustrate the idea, we shall
 first introduce the class of spectral mean that 
admits the form $G(f,\phi)=\int_0^{\pi}\phi(\lambda) f(\lambda)d\lambda$, where $f(\cdot)$ is the spectral density function of
$(X_t)_{t\in\Z}$ and $\phi: [-\pi,\pi]\rightarrow\R$  is a symmetric function with bounded variation. A sample analogue of $f(\lambda)$ is the periodogram,  $I_n(\lambda)=(2\pi n)^{-1}|\sum_{t=1}^{n}(X_t-\bar{X}_n)e^{it\lambda}|^2$ and a natural estimator of $G(f,\phi)$ is  $G(I_n,\phi)=\int_{0}^{\pi}\phi(\lambda) I_n(\lambda) d\lambda$. Often in practice, the quantity of interest is  the  normalized (ratio) version of $G(f,\phi)$, i.e. $R(f,\phi)=G(f,\phi)/G(f,{\bf 1})$, which is estimated by its sample counterpart  $R(I_n,\phi)$.
Prominent examples include $\phi(\lambda)=2\cos(m\lambda)$,
$m\in\N$ and $\phi(\lambda)={1}_{[0,x]}(\lambda)$, $x\in
[0,\pi]$. The former corresponds to
$G(f,\phi)={\gamma}(m)$
and
$R(f,\phi)={\rho}(m)$,  which have been covered by the framework in Section~\ref{sec:method}.
The latter corresponds to
$G(I_n,\phi)=F_n(x)=\int_{0}^{x}I_n(\lambda)d\lambda$ and
$R(I_n,\phi)=F_n(x)/F_n(\pi)$, which are
 ${n^{1/2}}$-consistent estimators of the spectral distribution function $G(f,\phi)=F(x)$ and its ratio counterpart $R(f,\phi)=F(x)/F(\pi)$.
Note that both $F(x)$ and $F(x)/F(\pi)$ are functionals of $F_{\infty}$.

Denote by $f_4(\cdot,\cdot,\cdot)$ the fourth-order cumulant spectral density of the process $X_t$. Under appropriate moment and weak dependence conditions (Brillinger 1969; Rosenblatt 1985; Dahlhaus 1985), we have
\[n^{1/2}\{G(I_n,\phi)-G(f,\phi)\}\rightarrow_{D} N\{0,\sigma^2(\phi)\},\]
where $\sigma^2(\phi)=2\pi\left\{\int_{0}^{\pi}\phi^2(\lambda)f^2(\lambda)d\lambda+\int_{0}^{\pi} \int_{0}^{\pi}\phi(w_1)\phi(w_2)f_4(w_1,-w_1,-w_2)dw_1dw_2 \right\}$.
 Confidence interval construction for $G(f,\phi)$ and $R(f,\phi)$ has been investigated by a few researchers.
 A standard approach is to find a consistent estimate of  $\sigma^2(\phi)$ and apply the plug-in principle, which inevitably involves the estimation of  the integral of the fourth order cumulant spectra (Taniguchi 1982; Keenan 1987; Chiu 1988). The procedures proposed all involve a choice of a smoothing parameter, for which no theoretical or empirical guidance has been given.
   Other works that avoid direct estimation  include Dahlhaus and Janas (1996) and Kreiss and Paparoditis (2003) on the frequency domain bootstrap method (Franke and H\"ardle 1992) and Nordman and Lahiri (2006) on the empirical-likelihood-based approach. A limitation of these  works is that their methodology heavily relies on the assumption that $X_t$ is a linear process with independent and identically distributed errors and may not be valid for more general stationary processes. Recently, Shao (2009) proposed a self-normalization-based approach that is widely applicable to a  large class of stationary process. It involves a  bandwidth, which is chosen by using information criteria and a moving average sieve approximation. Simulation studies show reasonably good finite sample performance. However,   a drawback of the method in Shao (2009) is that, for confidence intervals of $R(f,\phi)$,  there is a possibility that the method yields empty or no meaningful confidence intervals when the sample size is small. In contrast, the method that is developed in this article always delivers meaningful and nonempty intervals, and also there is no need to choose any tuning parameters in our procedure.

In what follows, we shall establish a theorem under a  general framework that includes the (normalized) spectral mean as a special case.  Suppose that the quantity of interest is $\theta=T(F_{\infty})\in \R^q$ and its estimator is $\hat{\theta}_n=T_n(\rho_n^1)$, where $T_n$ is a functional of the $n$-th dimensional distribution of $(X_t)_{t\in\Z}$ and it takes value in $\R^q$.  Denote by $\hat{\theta}_t=T_t(\rho_t^1)$ the estimator that is based on $(X_1,\cdots,X_t)$, $t=1,\cdots,n$.
The following theorem shows that it is possible to extend the validity of our method described in Section~\ref{sec:method} to a more general setting. The underlying idea is that if there are a sequence of approximating statistics for $\hat{\theta}_n$, that is a functional of the $B_n$-th dimensional empirical distribution ($B_n$ can be fixed or grows with $n$; see Remark~\ref{rem:Bn}), and a sequence of approximating quantities $\bar{\theta}_n$ for $\theta$,  then our method still delivers an (asymptotically) valid confidence interval provided that the approximation errors that are associated with  the approximating statistics and $\bar{\theta}_n$ are asymptotically negligible and similar regularity conditions hold for the expansion of the approximating statistics around $\bar{\theta}_n$. For the convenience of notation, let $r_n=\lfloor rn\rfloor$ for $r\in (0,1]$.

 \begin{theorem}
 \label{th:main2}

 Assume that there are a sequence of positive integers  $B_n$   and a sequence of  approximating quantities $\bar{\theta}_n$, that is a functional of $F_{B_n}$ and satisfies $|\theta-\bar{\theta}_n|=o(n^{-1/2})$. Let $Y_{tn}=(X_{t-B_n+1},\cdots,X_t)'$ and $\rho_{B_n}^{r_n}$ be the empirical distribution based on $(Y_{1n},\cdots,Y_{r_n n})'$ for $r\in (0,1]$. Further assume the expansion
 \[T_{B_n}(\rho_{B_n}^{r_n })-\bar{\theta}_n=r_n^{-1}\sum_{t=1}^{r_n} IF(Y_{tn};F_{B_n})+R_{r_n n}, \]
 where
 $\E\{IF(Y_{tn};F_{B_n})\}=0$,
 \begin{eqnarray}
 \label{eq:ip2}
 n^{-1/2}\sum_{t=1}^{r_n} IF(Y_{tn};F_{B_n})\Rightarrow \Delta B_q(r)
 \end{eqnarray}
 for some lower triangular matrix $\Delta$ with $\Delta\Delta'$ being positive definite.  Suppose that
 \begin{eqnarray}
 \label{eq:condition1}
 (i), ~R_{n n}=o_p(n^{-1/2})~ \mbox{and}~ n^{-2}\sum_{t=1}^{n} |tR_{tn}|^2 = o_p(1);
 \end{eqnarray}
 \begin{eqnarray}
 \label{eq:condition2}
 (ii), ~T_{B_n}(\rho_{B_n}^{n})-\hat{\theta}_{n}=o_p(n^{-1/2}) ~\mbox{and}~ n^{-2}\sum_{t=1}^{n} |t\{T_{B_n}(\rho_{B_n}^{t})-\hat{\theta}_{t}\}|^2=o_p(1).
 \end{eqnarray}
 Then $n(\hat{\theta}_n-\theta)'W_n^{-1} (\hat{\theta}_n-\theta)\rightarrow_{D} U_q$,
 where $W_n=n^{-2}\sum_{t=1}^{n}t^2(\hat{\theta}_t-\hat{\theta}_n)(\hat{\theta}_t-\hat{\theta}_n)'$.  Subsequently, the $100(1-\alpha)\%$ confidence region for
 $\theta$ is $\{\theta: n(\hat{\theta}_n-\theta)' W_n^{-1} (\hat{\theta}_n-\theta)\le U_{q,\alpha}\}$.
\end{theorem}

 \noindent Proof of Theorem~\ref{th:main2}: Let $\tilde{\theta}_{t}=T_{B_n}(\rho_{B_n}^{t})$. Following the argument  in the proof of Theorem
~\ref{th:main1}, we can show that  $n(\tilde{\theta}_n-\bar{\theta}_n)'\tilde{W}_n^{-1} (\tilde{\theta}_n-\bar{\theta}_n)'\rightarrow_{D} U_q$,
where $\tilde{W}_n=n^{-2}\sum_{t=1}^{n}t^2(\tilde{\theta}_t-\tilde{\theta}_n)(\tilde{\theta}_t-\tilde{\theta}_n)'$.
The conclusion follows from our assumption~(\ref{eq:condition2}) and $|\theta-\bar{\theta}_n|=o(n^{-1/2})$.

\qed

\begin{remark}
\label{rem:Bn}
{\rm If $B_n=m$ is fixed, then Theorem~\ref{th:main2} generalizes Theorem~\ref{th:main1} by allowing $\bar{\theta}_n$ to be dependent on $n$, and $\hat{\theta}_n$ to be slightly different from the approximately linear statistic that was defined before. For example, if the quantity of interest is $\gamma(m-1)$, then Theorem~\ref{th:main1} is only applicable to the statistic $\tilde{\gamma}_n(m-1)$, not to $\hat{\gamma}_n(m-1)$. With the formulation of Theorem~\ref{th:main2}, we can let $\hat{\theta}_{r_n}=\hat{\gamma}_{r_n}(m-1)$ for $r_n>m$, $T_{B_n}(\rho_{B_n}^{r_n})=T_m(\rho_m^{r_n})=\tilde{\gamma}_{r_n}(m-1)$ and $\bar{\theta}_n=\theta=\gamma(m-1)$. It is easy to verify that the technical assumptions in Theorem~\ref{th:main2} hold under mild moment and weakly dependent conditions on $X_t$. The details are omitted. If $B_n\rightarrow\infty$ as $n\rightarrow\infty$, then we typically require $B_n/n\rightarrow 0$ as shown in the example of spectral mean below.
}
\end{remark}

To illustrate the verifiability of the assumptions in Theorem~\ref{th:main2}, we focus on  the case for the spectral mean.
For simplicity, we assume that $\E(X_t)=0$ is known. The complication that is caused by the mean correction can be handled with additional routine technical details.  Letting $\psi_k=(2\pi)^{-1}\int_0^{\pi}\phi(\lambda)e^{ik\lambda}d\lambda$,
  then $G(I_n,\phi)=\sum_{k=1-n}^{n-1}\hat{\gamma}_n(k)\psi_k=\sum_{k=0}^{n-1}\hat{\gamma}_n(k)g_k$, with $g_k=(\psi_k+\psi_{-k})$ if $k\not=0$, and $g_0=\psi_0$. Similarly, we have
  $G(f,\phi)=\sum_{k=0}^{\infty}\gamma(k) g_k$.
   Let $\bar{\theta}_n=\sum_{k=0}^{B_n-1}\gamma(k) g_k$. Denote by $\hat{\gamma}_{r_n}(k)=r_n^{-1}\sum_{t=1}^{r_n-|k|}X_tX_{t+|k|}$. Then
    $T_{B_n}(\rho_{B_n}^{r_n})=\sum_{k=0}^{B_n-1}\hat{\gamma}_{r_n}(k) g_k$,
      $IF(Y_{tn};F_{B_n})=\sum_{k=0}^{B_n-1}\{X_tX_{t-k}-\gamma(k)\} g_k$,
      $\hat{\theta}_{r_n}=\sum_{k=0}^{r_n -1} \hat{\gamma}_{r_n}(k) g_k$ and $R_{r_n n}=-r_n^{-1}\sum_{k=0}^{B_n-1}\sum_{t=1}^{k} X_tX_{t-k} g_k$.

  In the case of the spectral mean, the functional central limit theorem (\ref{eq:ip2}) has been established in Theorem 1 of Shao (2009). The following proposition shows that the other two key assumptions (\ref{eq:condition1}) and (\ref{eq:condition2}) in
      Theorem~\ref{th:main2} can be verified as well.

      \begin{proposition}
    \label{prop:1}
      Assume that $1/B_n+B_n/n=o(1)$, $\sum_{j=1}^{\infty}g_j^2<\infty$ and $\sum_{k=B_n}^{\infty}|\gamma(k)|=o(n^{-1/2})$. Further assume that
      \begin{eqnarray}
      \label{eq:summability}
      \sum_{k\in\Z}|\gamma(k)|<\infty~~\mbox{and}~\sum_{k_1,k_2,k_3\in\Z}|\cum(X_0,X_{k_1},X_{k_2},X_{k_3})|<\infty.
      \end{eqnarray}
      Then $r_n^2\E|T_{B_n}(\rho_{B_n}^{r_n})-\hat{\theta}_{r_n}|^2=o(n)$ and $r_n^2\E|R_{r_n n}|^2=o(n)$ uniformly in $r\in (0,1]$.
      \end{proposition}

It is straightforward to see that the assumptions~(\ref{eq:condition1}) and (\ref{eq:condition2}) follow from the conclusion of Proposition~\ref{prop:1}.
 Note that  $|\theta-\bar{\theta}_n|=o(n^{-1/2})$ if $\sum_{k=B_n}^{\infty}|\gamma(k)|=o(n^{-1/2})$. Hence, the assumptions in Theorem~\ref{th:main2}
 are all satisfied for the spectral mean. The case for the normalized spectral mean can be treated in a similar fashion. We omit the details.

\section{Finite sample performance}
\label{sec:sim}

 Through Monte Carlo simulations, we investigate the size and power properties of the test statistic $\tilde{T}_K$ in Section~\ref{subsec:1},  the empirical coverage probabilities of confidence intervals for (normalized) spectral means, the median and unknown parameter vector in time series models by using $M$-estimation in Section~\ref{subsec:2}.  We also compare the self-normalized approach with the standard and bootstrap approaches via simulations in Section~\ref{subsec:4}.

\subsection{Size and power of $\tilde{T}_K$ statistic}
\label{subsec:1}

We first investigate the size of $\tilde{T}_K$ and compare it with  $T_K$'s of Lobato (2001) and $\tilde{Q}_K$'s of Lobato et al. (2002) at $K=1,3,5$. Only the case $K=1$ is examined for $T_K$ in Lobato (2001).  For the $\tilde{Q}_K$ test, it corresponds to efficient Studentization, so a bandwidth parameter is involved in the consistent estimation of the asymptotic covariance matrix of the first $K$ sample correlations. Here we adopt an automatic procedure as used in Lobato et al. (2002), i.e., we employ the AR(1) prewhitening and selects the bandwidth by using formula (2.2) of Newey and West (1994) with weights equal to one and lag truncation equal to $2(n/100)^{2/9}$. Let $u_t$ stand for a sequence of iid standard normal random variables.
For the comparison, we use the same models as studied in Lobato (2001). They are (1), iid N(0,1); (2), iid t$(6)$;
 (3),  demeaned standard log normal; (4), the 1-dependent process $X_t=u_tu_{t-1}$; (5), the heteroscedastic process $X_t=s_tu_tu_{t-1}$, where $s_t$ is the infinite repetition    of the sequence $\{1,1,1,2,3,1,1,1,1,2,4,6\}$; (6), the uncorrelated non-martingale difference process $X_t=u_{t-2}u_{t-1}(u_{t-2}+u_t+1)$;
    (7), the GARCH$(1,1)$ process $X_t=u_t\sigma_t$, where $\sigma_t^2=0.001+0.02X_{t-1}^2+0.8\sigma_{t-1}^2$; (8), the bilinear model $X_t=u_t+0.5u_{t-1}X_{t-2}$.

Two sample sizes $n=100$ and $n=500$ are investigated with 5000 replications. As seen from Table~\ref{tb:table1}, the size distortion increases as $K$ increases, and it improves as we enlarge the sample size with the improvement almost being uniform over all the models and methods. The test statistic $\tilde{T}_K$ tends to produce a higher size than $T_K$. When $K=3$ and $K=5$,  $T_K$ is substantially under-sized for models (3)-(6) at $n=100$, whereas the size distortion is noticeably less for $\tilde{T}_K$. In contrast, for models (1), (7) and (8), $\tilde{T}_K$ is outperformed by $T_K$ in terms of size distortion. A comparison of the size for $\tilde{Q}_K$ with that of $T_K$ and $\tilde{T}_K$ shows that the size performance of $\tilde{Q}_K$ is less satisfactory. The  $\tilde{Q}_K$  test tends to be undersized and its size distortion appears to be very severe
  for some models (e.g. models (5) and (6)) at $K=3$ and $K=5$ even when $n=500$. Since  the size distortion is closely related to the bandwidth selection algorithm (Newey and West 1994), this seems to suggest that the particular data-driven bandwidth that we used here does not perform uniformly well across different models for a large sample size.

  \centerline{Please insert Table~\ref{tb:table1} here!}

To investigate the power, we reconsider the models that were used in Lobato (2001), i.e.,  the AR$(1)$ model with innovations following both the GARCH$(1,1)$ process that was specified in (7) above and the bilinear models in (8). The  autoregressive coefficient $\rho$ varies from 0.1 to 0.5 with a spacing of 0.1. The sample size is taken to be $n=100$ and the number of replications is 5000. Table~\ref{tb:table2} shows the size-adjusted empirical rejection percentages for $K=1,3,5$ at $5\%$ and $10\%$ levels. In general, the larger $K$ is, the lower the rejection rate becomes. For both models, the power of $\tilde{T}_K$ is fairly close to that for $T_K$.  As for $\tilde{Q}_K$, its power advantage over $T_K$ and $\tilde{T}_K$ is pronounced when $\rho=0.2,0.3,0.4$, but seems to diminish as $K$ increases from $K=1$ to $K=3$ and $K=5$. When $\rho=0.5$, $K=3$ or $K=5$, the power of $\tilde{Q}_K$ is close to or even slightly worse than that of $T_K$ and $\tilde{T}_K$ in some cases, which suggests some theoretical investigation. Overall, it seems fair to conclude that $\tilde{Q}_K$ has moderately more power but worse size than $T_K$ and $\tilde{T}_K$, for which the size and power properties are comparable. This observation is consistent with the ``better size but less power" of the self-normalized approach  compared with its efficiently Studentized counterpart.

   \centerline{Please insert Table~\ref{tb:table2} here!}

\subsection{Spectral means, Median and M-estimators}
\label{subsec:2}

In this subsection, we first examine the coverage probability for spectral means $\gamma(1)$ and $F(\pi/2)$ as well as
their ratio counterparts $\rho(1)$ and $F(\pi/2)/F(\pi)$. Let  $B$ be  the backward shift operator, and $\varepsilon_{1t}$ and $\varepsilon_{2t}$ be iid   with  $N(0,1)$ and $t(5)$ distributions respectively,  $\varepsilon_{3t}=u_t\{{0.5\varepsilon_{3(t-1)}^2+0.3}\}^{1/2}$ follows an ARCH$(1)$ process.  We follow the setup in Shao (2009) and consider two sample sizes $n=150$ and $n=600$ and the following six models:

\noindent $M_1$: $(1-0.7B)X_t=\varepsilon_{1t}$; $M_2$: $(1-0.7B)X_{t}=0.6^{1/2} \varepsilon_{2t}$;
$M_3$: $(1-0.7B)X_{t}=\varepsilon_{3t}/{0.6}^{1/2}$; $M_4$: $X_t=(1+0.8B)\varepsilon_{1t}$;  $M_5$: $X_t=(1+0.8B)\{{0.6}^{1/2}\varepsilon_{2t}\}$;
 $M_6$: $X_t=(1+0.8B)\{\varepsilon_{3t}/{0.6}^{1/2}\}$.
In these models,  the variances for $t(5)$ and ARCH$(1)$ processes are standardized to $1$. The number of replications is 1000.
 As seen from Tables~\ref{tb:table3} and \ref{tb:table4},   the coverages offered by our method are comparable with those delivered  by Shao's (2009) approach for all models and sample sizes under consideration. The coverages of the intervals for spectral mean (e.g. $\gamma(1)$)  are farther from  the nominal level than those for their ratio counterparts (e.g. $\rho(1)$), which is consistent with the finding in Shao (2009).
 For both $\rho(1)$ and $F(\pi/2)/F(\pi)$,   a portion of intervals are empty   at $n=150$ using the method in Shao (2009), whereas our approach always produces an $[L,U]$-type non-empty interval.  It is also worth noting that the coverages of the new method for models with ARCH errors are close to the nominal level when $n=600$, suggesting that it is applicable to linear processes with dependent innovations. Since the new method is bandwidth free, has comparably good finite sample coverage and wide applicability, it seems preferable to Shao (2009).

 As suggested by a referee, we also include the coverage percentages for $\gamma(1)$ and $\rho(1)$ using the efficiently Studentized approach, which involves consistent estimation of the asymptotic variances of $\hat{\gamma}_n(1)$ and $\hat{\rho}_n(1)$. In particular, we follow the idea that was presented in  Lobato et al. (2002) and slightly modify their procedure. Let $\hat{\gamma}=(\hat{\gamma}_n(0),\hat{\gamma}_n(1))'$, $\gamma=(\gamma(0),\gamma(1))'$,  $\hat{w}_{0t}=(X_t-\bar{X}_n)(X_{t}-\bar{X}_n)$ and   $\hat{w}_{1t}=(X_t-\bar{X}_n)(X_{t-1}-\bar{X}_n),~t=2,\cdots,n$.
 Under suitable conditions, we have $\sqrt{n}(\hat{\gamma}-\gamma)\rightarrow_{D} N(0,V)$, where $V$ is a $2\times 2$ matrix with elements $(V_{00},V_{01}; V_{10}, V_{11})$. We estimate $V$ by applying the lag window method to $\hat{w}_{t-1}=(\hat{w}_{0t},\hat{w}_{1t})'$ for $t=2,\cdots,n$, i.e.,
 $\hat{V}=(n-1)^{-1}\sum_{j}\sum_{t}K(j/l)(\hat{w}_t-\bar{\hat{w}})(\hat{w}_{t-j}-\bar{\hat{w}})'$,
 where $\bar{\hat{w}}=(n-1)^{-1}\sum_{t=1}^{n-1}\hat{w}_{t}$. We use the Bartlett kernel for $K$ and the same bandwidth selection algorithm (Newey and West 1994) as adopted in $\tilde{Q}_K$ test. To estimate the asymptotic variance of $\sqrt{n}\{\hat{\rho}_n(1)-\rho(1)\}$, we plug in the estimates for the unknown quantities in equation (1) of Lobato et al. (2002), and the resulting estimate is
 $\hat{\gamma}_n(0)^{-2}[\hat{V}_{11}-\hat{\rho}_n(1)\hat{V}_{10}-\hat{\rho}_n(1)\hat{V}_{01}+\hat{\rho}_n(1)^2\hat{V}_{00}]$.
 As seen from Table~\ref{tb:table3},  the efficiently Studentized approach exhibits undercoverage in the case of $\gamma(1)$ and its coverage is noticeably worse than those of the other two methods for all the models and sample sizes under consideration. For $\rho(1)$, the efficiently Studentized approach delivers reasonably good
coverage, with apparent over-coverage for models with $t(5)$ errors; see the results for $M_2$ and $M_5$. It is not clear why the coverage gets worse for some models (e.g., $M_2$, $M_4$, $M_5$ ) when the sample size increases from $150$ to $600$. Nevertheless, the coverage performance of the self-normalized approach is at least not inferior to the efficiently Studentized approach.

 \centerline{Please insert Tables~\ref{tb:table3}\&\ref{tb:table4} here!}

In what follows, we further examine the coverage accuracy for $med(X_1)$ on the basis of models $M_1$-$M_6$, and for the unknown parameter vector in the models  $M_1$-$M_3$ and the following three AR$(2)$ models using the least absolute deviation (LAD) estimates. Let $M_7$: $(1-\phi_1 B-\phi_2 B^2)X_t=\varepsilon_{1t}$;  $M_8$: $(1-\phi_1 B-\phi_2 B^2)X_t=0.6^{1/2} \varepsilon_{2t}$;  $M_9$: $(1-\phi_1 B-\phi_2 B^2)X_t=\varepsilon_{3t}/{0.6}^{1/2}$. The true value of $(\phi_1,\phi_2)=(0.6,0.35)$. For models $M_7$-$M_9$,  we estimate $(\phi_1,\phi_2)$ by  $(\hat{\phi}_{1n},\hat{\phi}_{2n})=\mbox{argmin}_{(\phi_1,\phi_2)}\sum_{t=3}^{n}|X_t-\phi_1 X_{t-1}-\phi_2X_{t-2}|$.  Table~(\ref{tb:table5}a) shows that in the case of  the median, there is undercoverage for all the models and sample sizes. The coverage is fairly close to the nominal level when $n=600$.
In addition, the difference in the models' innovation distribution does not seem to affect the coverage much. For the LAD estimates, overcoverage occurs for both sample sizes and all models, and there seems more distortion at the $90\%$ level than at the $95\%$ level.   From Table~(\ref{tb:table5}b), the overcoverage appears  more severe for models $M_7$-$M_9$ than for models $M_1$-$M_3$, which is because the asymptotic approximation tends to become  worse when $q$ (i.e. the number of unknown parameters) gets larger.

 \centerline{Please insert Table~\ref{tb:table5} here!}

\subsection{Block bootstrap, normal approximation and self-normalization}
\label{subsec:4}

The focus of this subsection is to compare the finite sample coverages of the self-normalized approach with the standard approach, where consistent
estimation of the asymptotic variance matrix is involved.   We consider the AR$(1)$ model, $X_t=\rho X_{t-1}+u_t$, where $u_t\sim $ iid $N(0,1)$ and $\rho=0,0.5$ and $0.8$. The sample size $n=50$ and the number of bootstrap replications is $1000$. We examine the empirical coverages of confidence intervals for $\E(X_1)$, $med(X_1)$ and  $\rho(1)$ based on 2000 replications and for $F(\pi/2)/F(\pi)$ based on 500 replications.  For the linear regression models with auto-correlated errors,  Goncalves and Vogelsang (2008) showed that the conventional block bootstrap test, where the formula for the bootstrapped standard error  admits the same form as that used on the original data, could be more accurate than the standard normal approximation under the fixed-b asymptotic framework. Their simulation results also indicate that, when the block size is suitably chosen, it may outperform the fixed-b approximation. In light of the findings in Goncalves and Vogelsang (2008) in the testing context, we shall incorporate  the block bootstrap method into our simulation studies. Let $\{X_t^*\}_{t=1}^{n}$ denote the bootstrap sample with block size $l=1,\cdots,15$. If $n/l$ is not an integer, we use a fraction of the last sampled block.  We compare the empirical coverages of the following four schemes at the  $95\%$ nominal level: (1), Moving block bootstrap without Studentizing. In other words, we approximate the sampling distribution of $\sqrt{N}(\hat{\theta}_N-\theta)$ by
$\sqrt{N}(\hat{\theta}_N^*-\hat{\theta}_N)$, where $\hat{\theta}_N^*$ is the functional $T$ applied to the bootstrap sample. So the 95\% confidence interval of $\theta$ is $[\hat{\theta}_N-q_{N,0.975}^{*}/\sqrt{N},\hat{\theta}_N-q_{N,0.025}^*/\sqrt{N}]$,
 where $q_{N,\alpha}^*$ denotes the $100\alpha$\%
percentile of $\sqrt{N}(\hat{\theta}_N^*-\hat{\theta}_N)$ based on $1000$ bootstrap replicates. (2), Normal approximation. To use standard normal approximation, we need a consistent estimate for the asymptotic variance. Here we use the block bootstrap variance estimator, which is denoted as $\hat{\sigma}^2$, whose consistency has been shown in K\"unsch (1989) and B\"uhlmann and K\"unsch (1995) for a large class of approximately linear statistics.
 The resulting confidence interval for $\theta$ is $[\hat{\theta}_N-1.96\hat{\sigma}/\sqrt{N},\hat{\theta}_N+1.96\hat{\sigma}/\sqrt{N}]$.
Note that other types of bandwidth-dependent consistent estimates are available, but it requires a case-by-case study. In contrast, the use of the block bootstrap method   allows us to treat the consistent estimation of asymptotic variances  for all the cases in a unified way. In addition, the implementation is very straightforward. (3), Moving block bootstrap with inefficient Studentizing. In this scheme, we approximate the sampling distribution of $N(\hat{\theta}_N-\theta)'W_N^{-1}(\hat{\theta}_N-\theta)$ by its bootstrap counterpart $N(\hat{\theta}_N^*-\hat{\theta}_N)'(W_N^*)^{-1}(\hat{\theta}_N^*-\hat{\theta}_N)$, where $W_N^*$ is obtained by plugging  the bootstrap sample into $W_N$. The 95\% confidence interval for $\theta$ is obtained by solving
\[N(\hat{\theta}_N-\theta)'W_N^{-1}(\hat{\theta}_N-\theta)\le U_{q,0.05}^*,\]
where $U_{q,0.05}^*$ stands for the $95\%$ percentile  of $N(\hat{\theta}_N^*-\hat{\theta}_N)'(W_N^*)^{-1}(\hat{\theta}_N^*-\hat{\theta}_N)$ based on 1000 bootstrap replicates. (4), The self-normalization-based approximation; compare Theorems~\ref{th:main1} and \ref{th:main2}.

\begin{figure}
\caption{Empirical coverage probabilities (left panel) and  ratios of the interval widths over that delivered by
 the self-normalized method (right panel) for $\E(X_1)$. Sample size $n=50$ and number of replications is 2000. }
\begin{center}
{\includegraphics[height=8cm,width=6cm,angle=270]{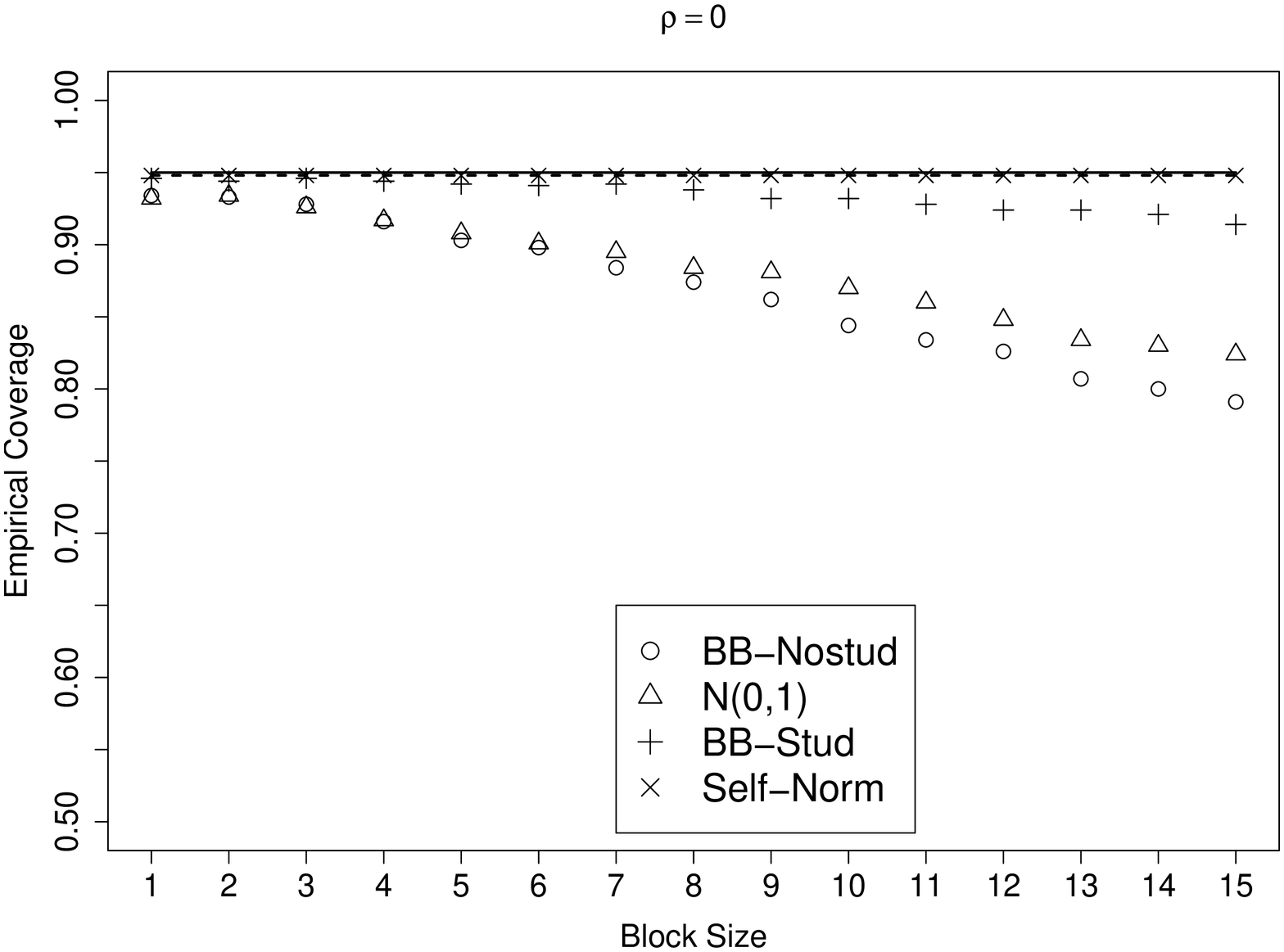}}
{\includegraphics[height=8cm,width=6cm,angle=270]{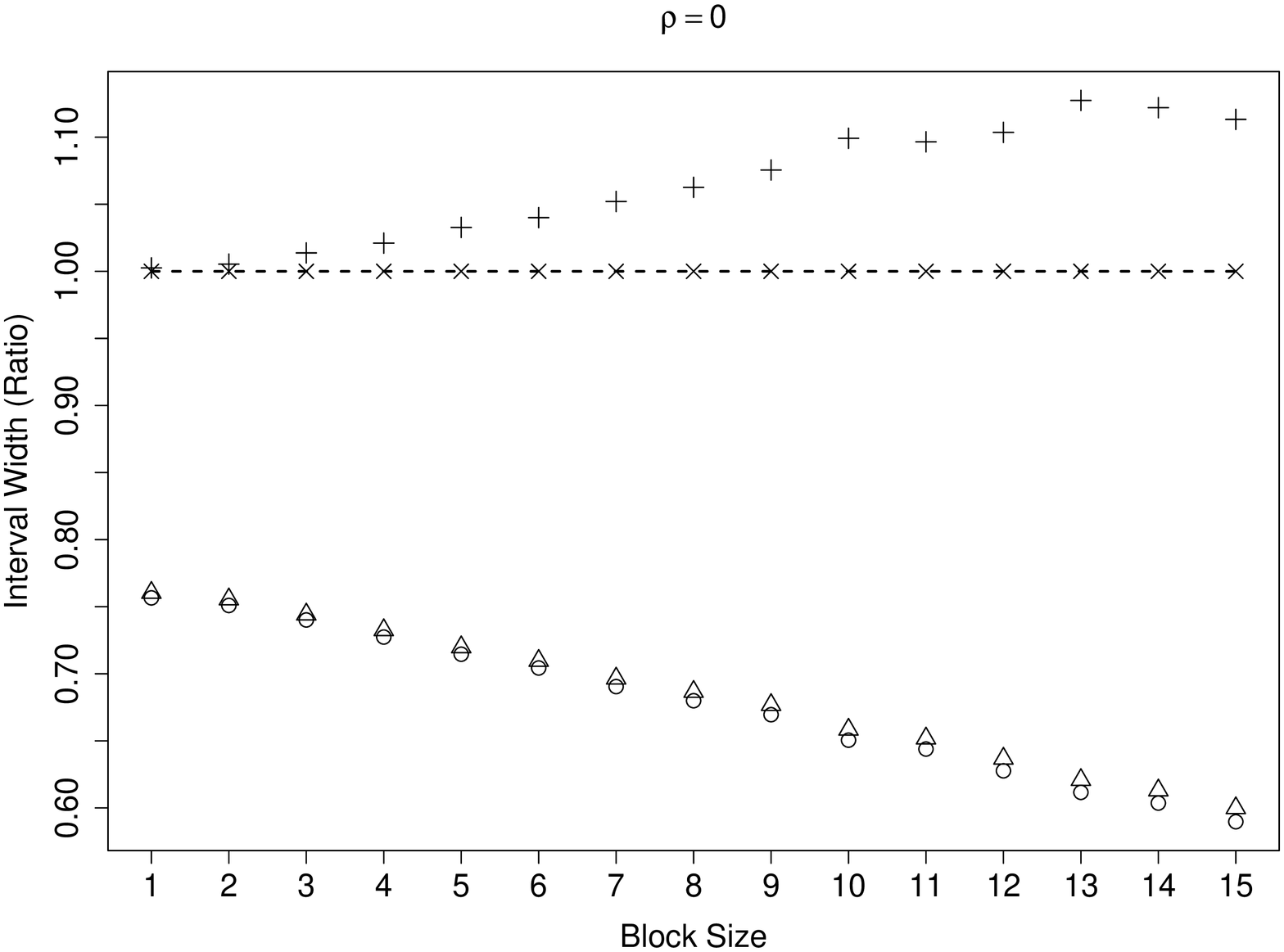}}
{\includegraphics[height=8cm,width=6cm,angle=270]{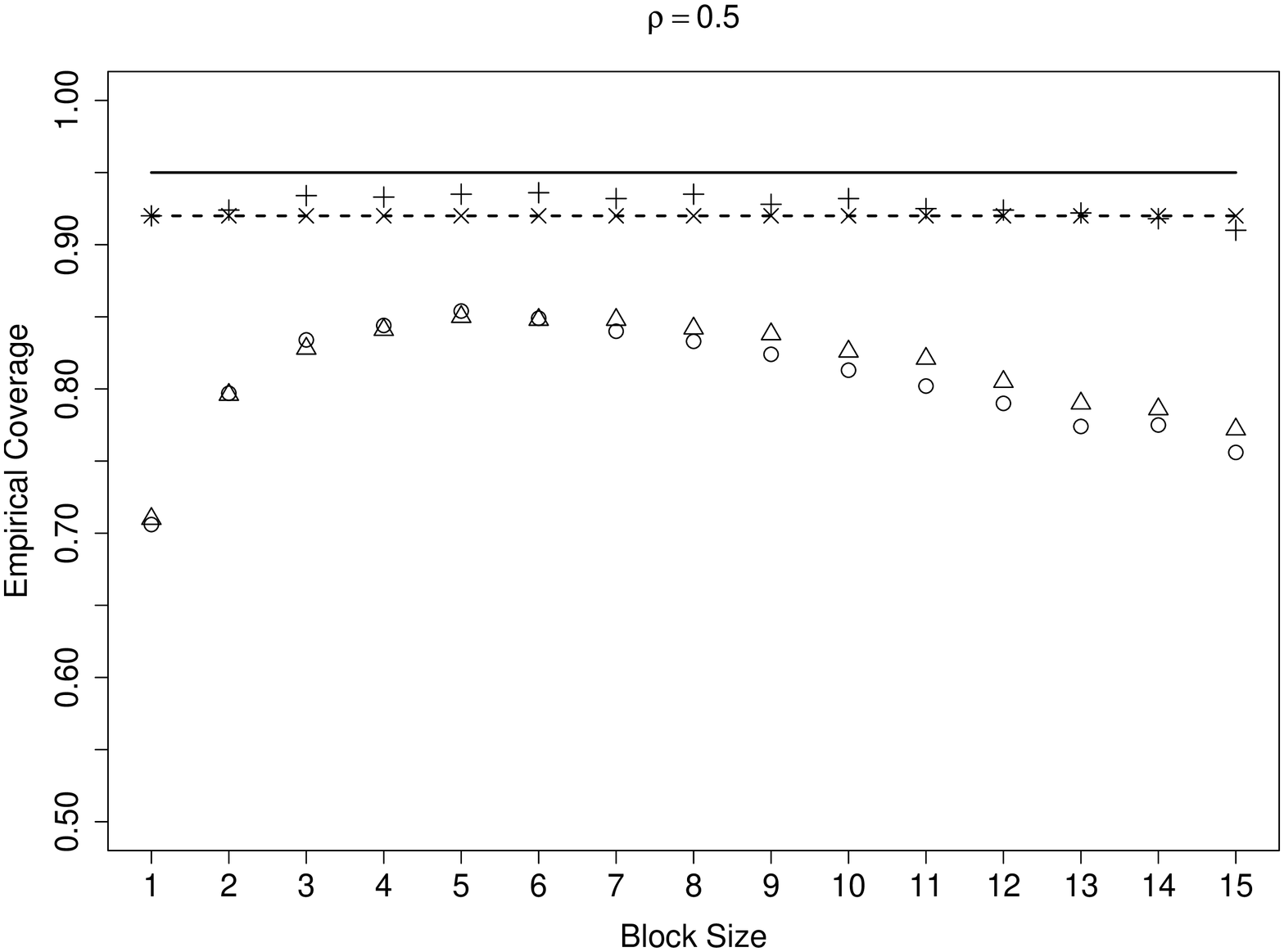}}
{\includegraphics[height=8cm,width=6cm,angle=270]{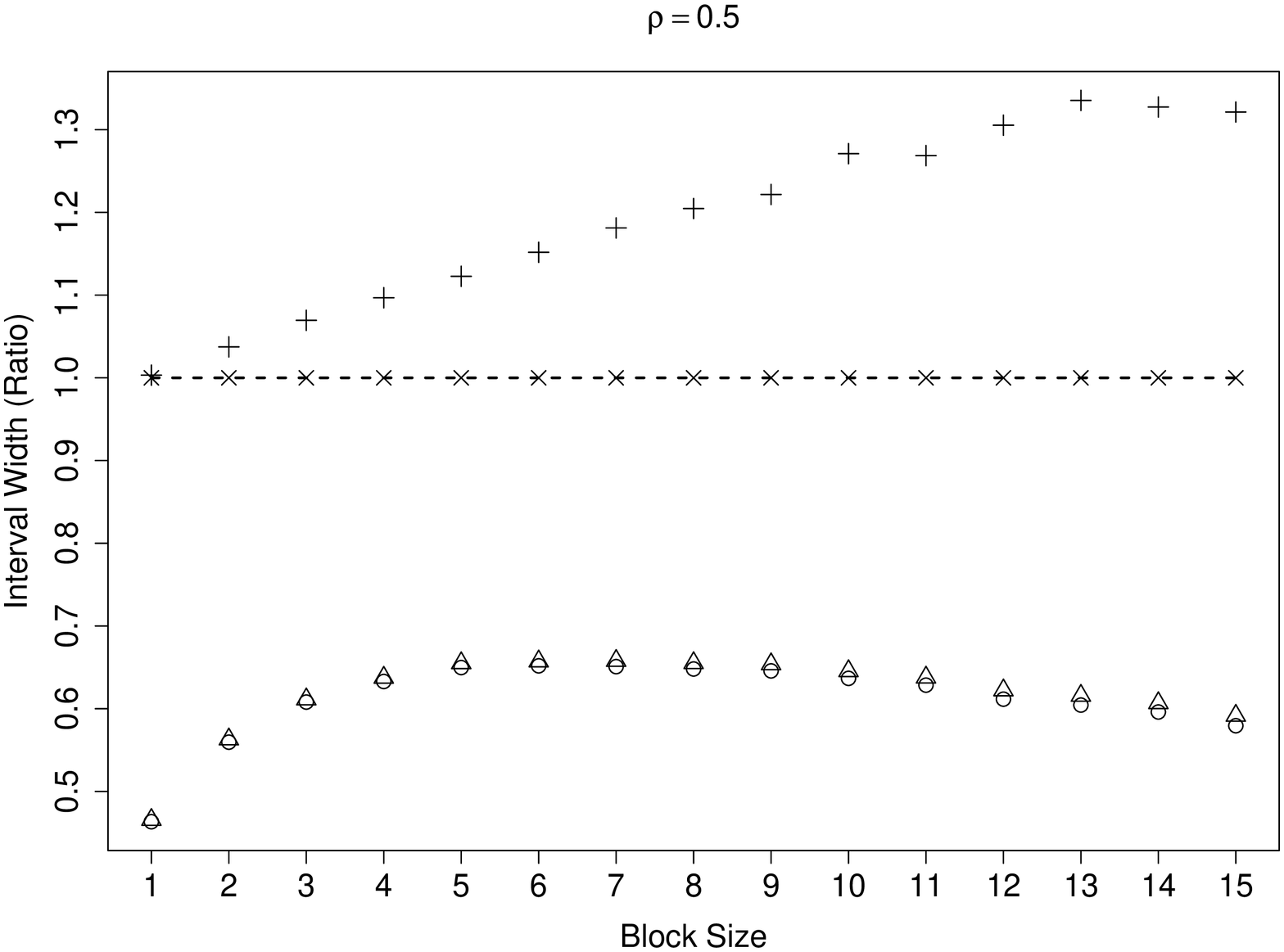}}
{\includegraphics[height=8cm,width=6cm,angle=270]{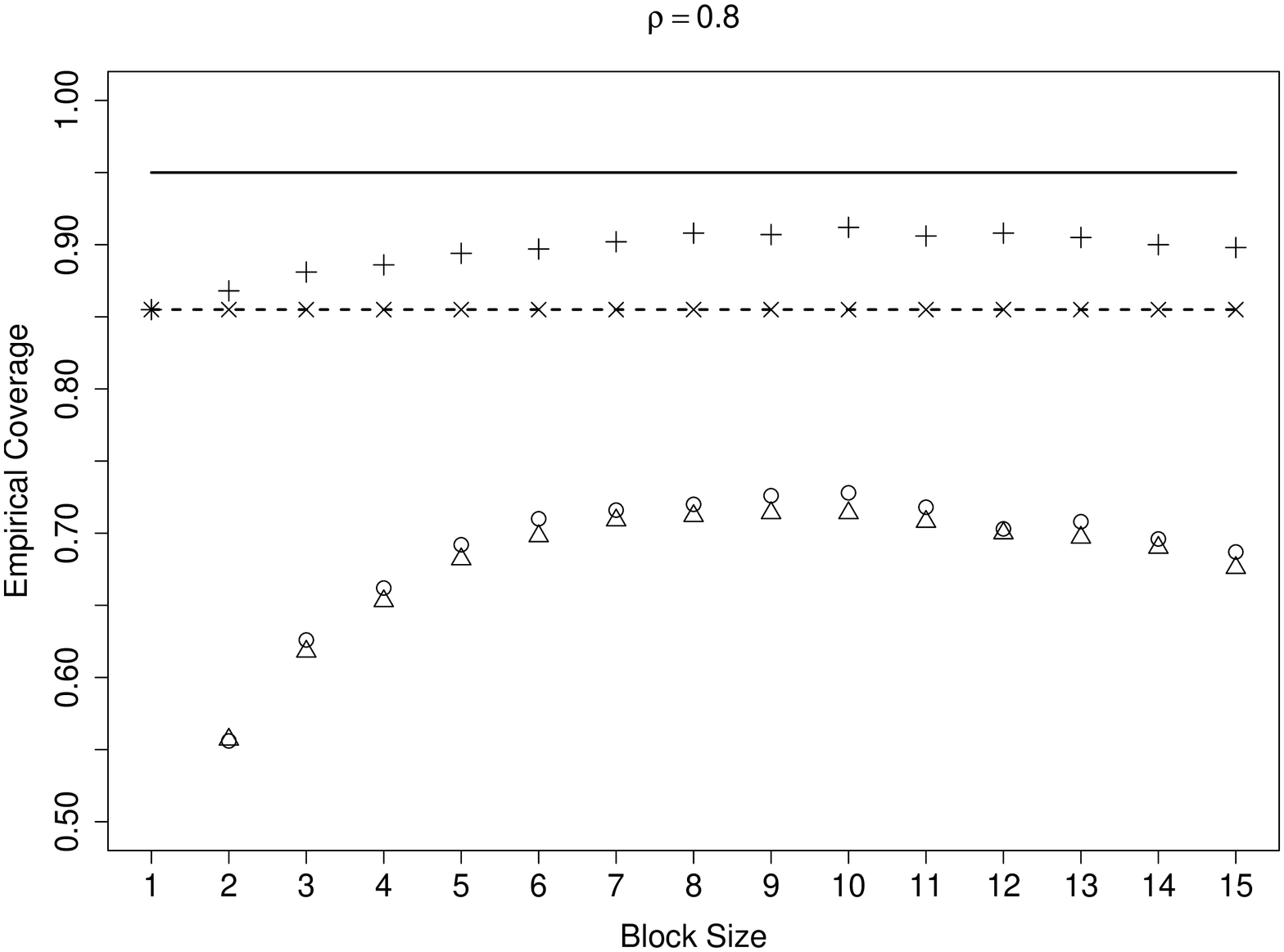}}
{\includegraphics[height=8cm,width=6cm,angle=270]{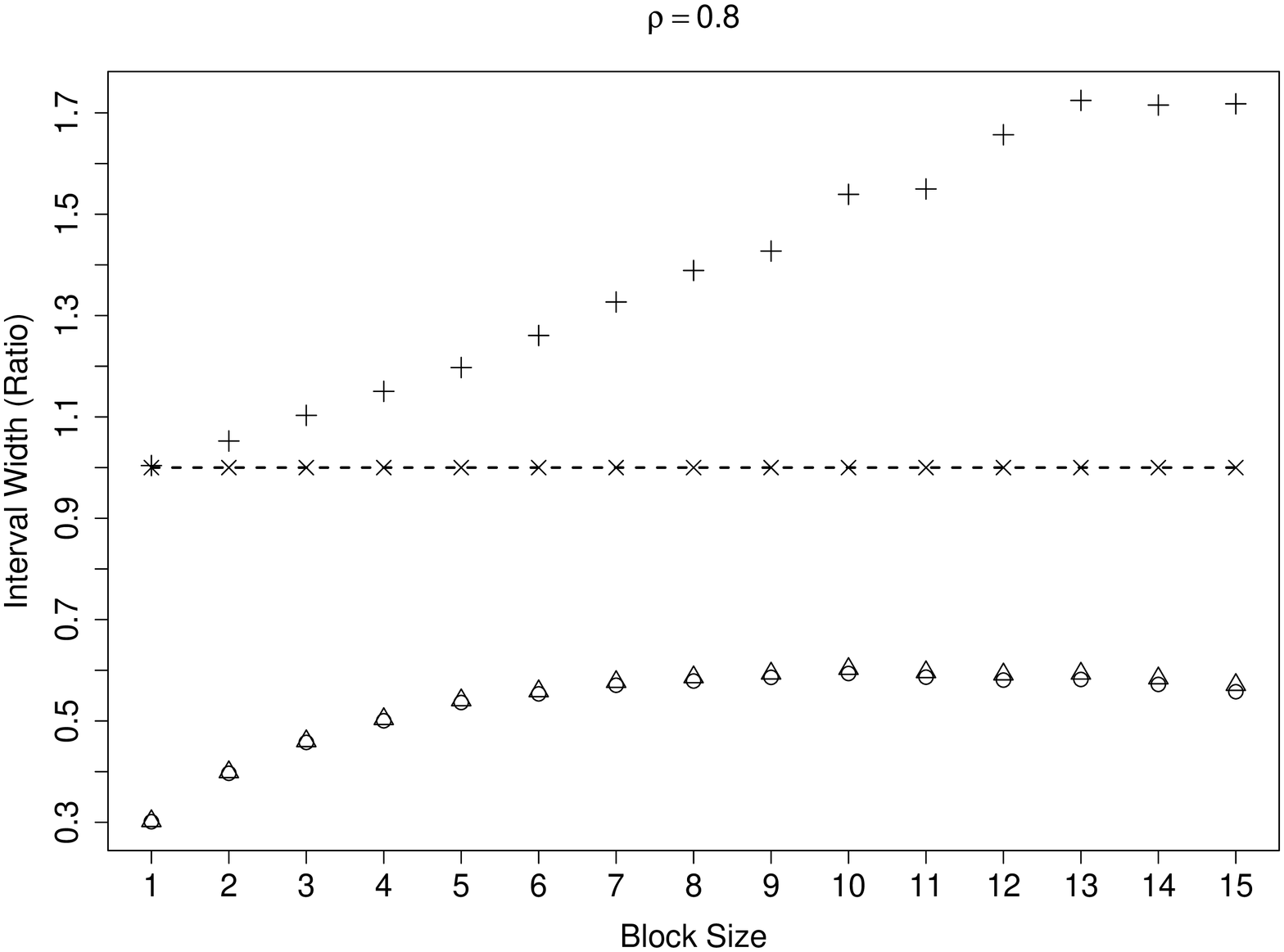}}
\label{fig:mean}
\end{center}

\end{figure}

\begin{figure}
\caption{Empirical coverage probabilities (left panel) and  ratios of the interval widths over that delivered by
 the self-normalized method (right panel) for $med(X_1)$. Sample size $n=50$ and number of replications is 2000. }
\begin{center}
{\includegraphics[height=8cm,width=6cm,angle=270]{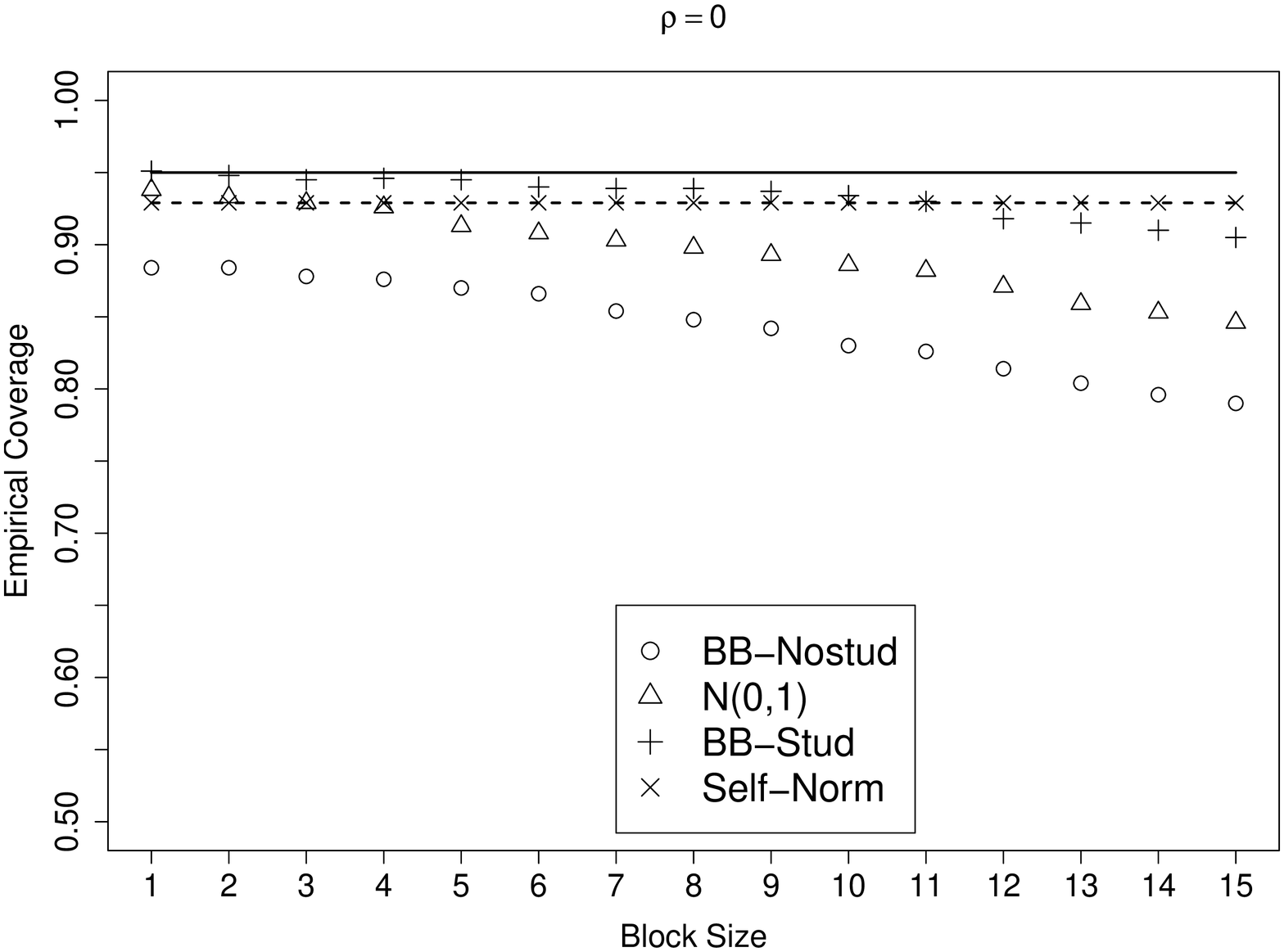}}
{\includegraphics[height=8cm,width=6cm,angle=270]{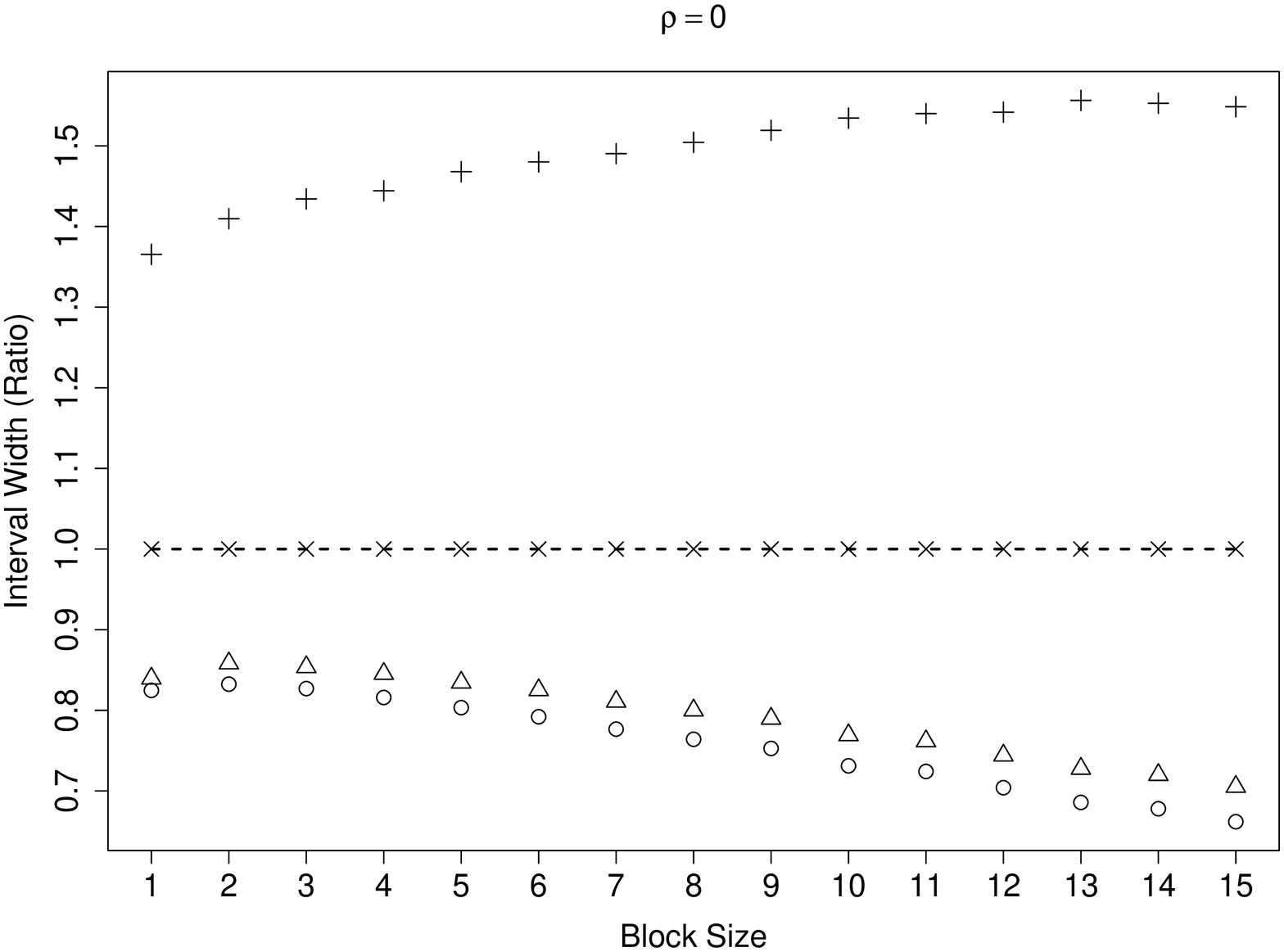}}
{\includegraphics[height=8cm,width=6cm,angle=270]{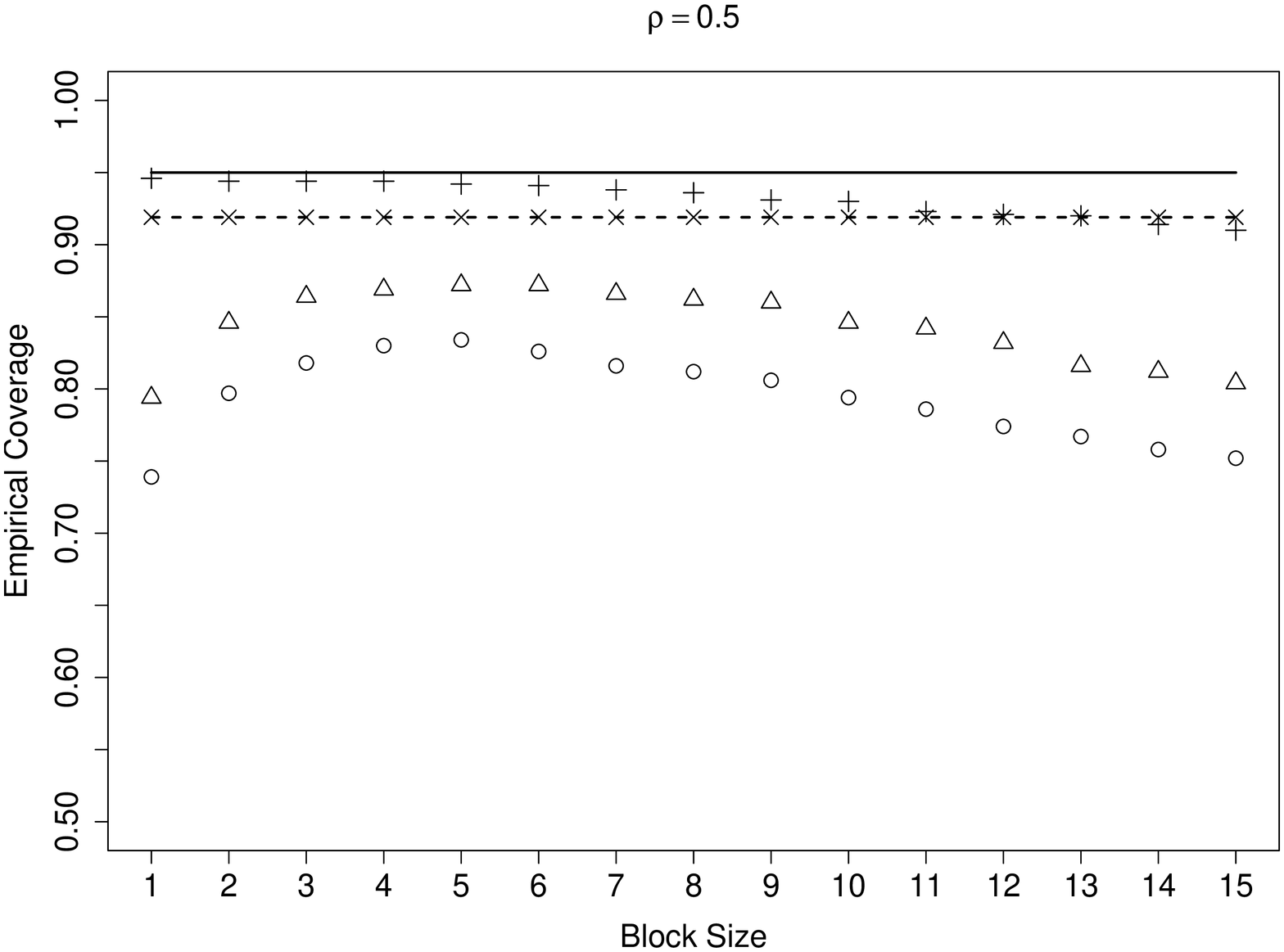}}
{\includegraphics[height=8cm,width=6cm,angle=270]{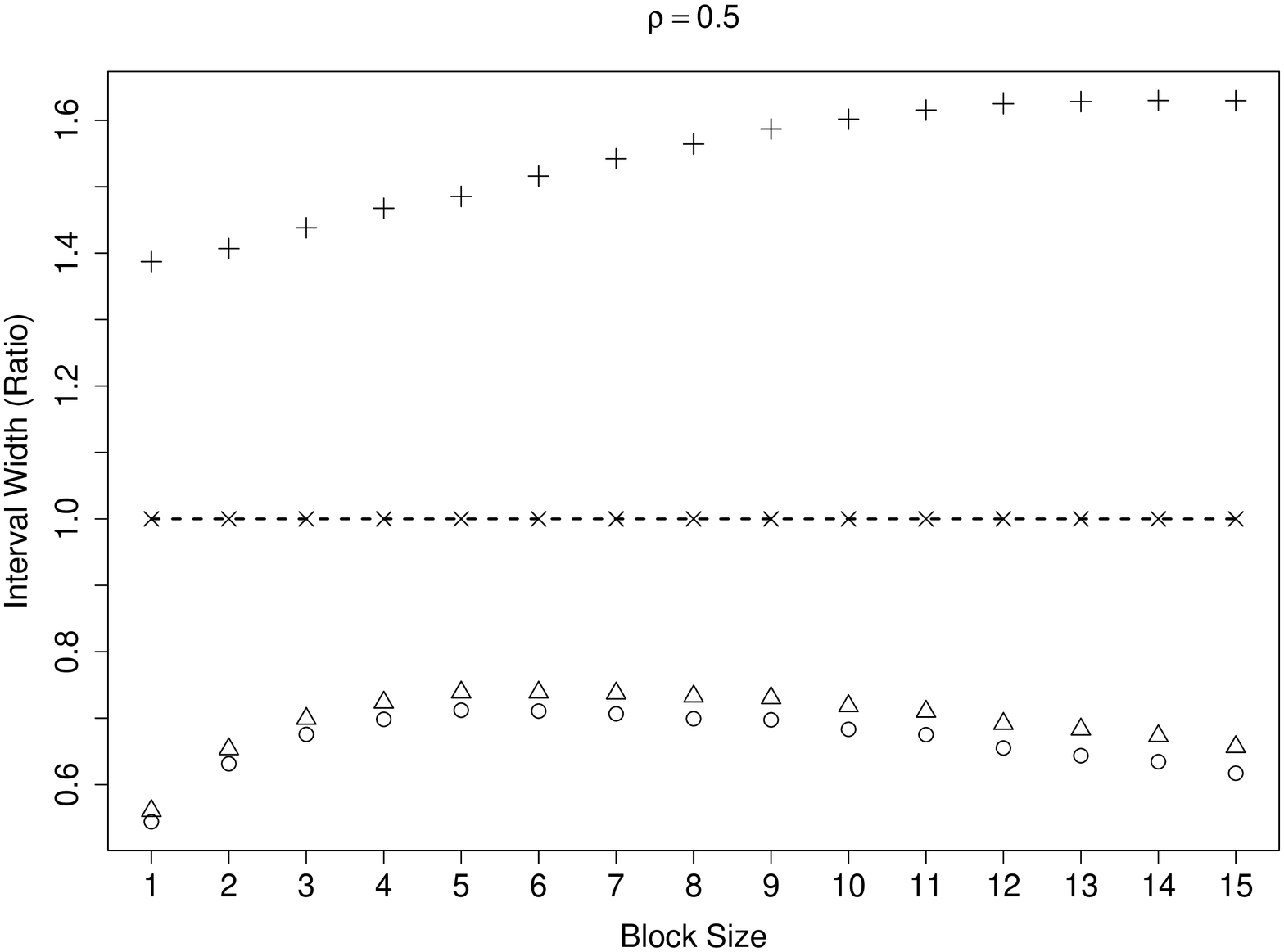}}
{\includegraphics[height=8cm,width=6cm,angle=270]{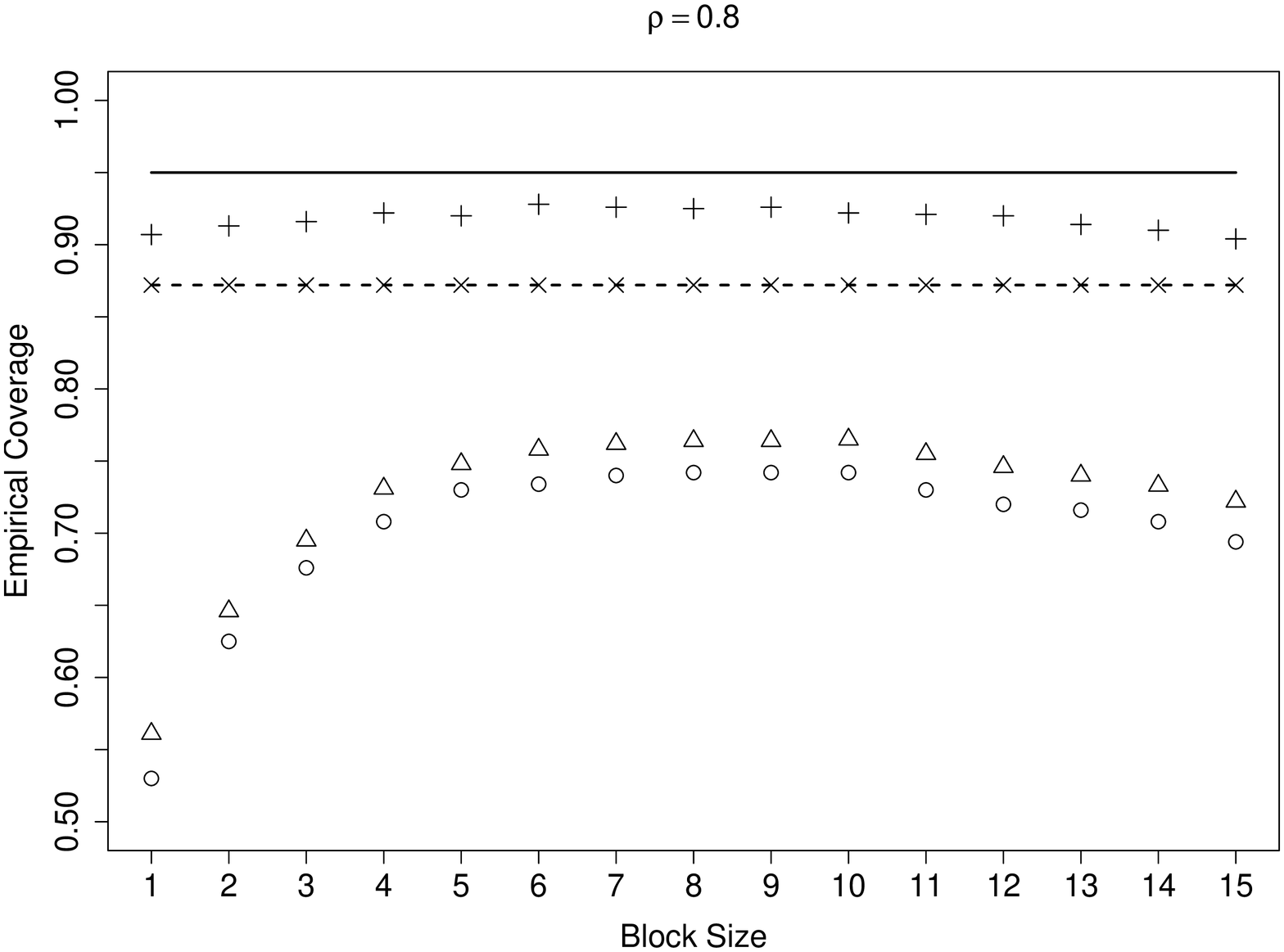}}
{\includegraphics[height=8cm,width=6cm,angle=270]{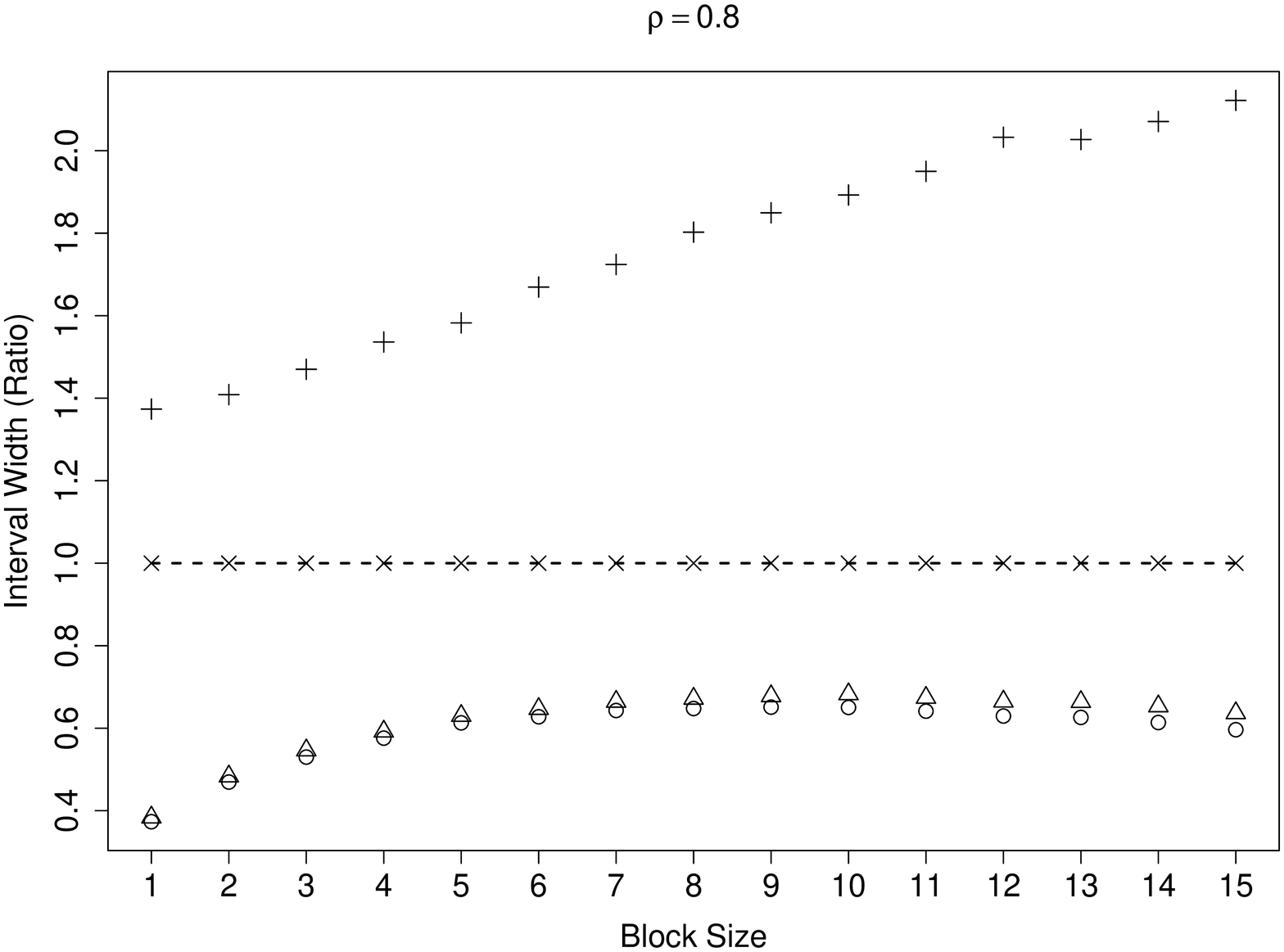}}
\label{fig:med}
\end{center}

\end{figure}

\begin{figure}
\caption{Empirical coverage probabilities (left panel) and  ratios of the interval widths over that delivered by
 the self-normalized method (right panel) for $\rho(1)$. Sample size $n=50$ and number of replications is 2000. }
\begin{center}
{\includegraphics[height=8cm,width=6cm,angle=270]{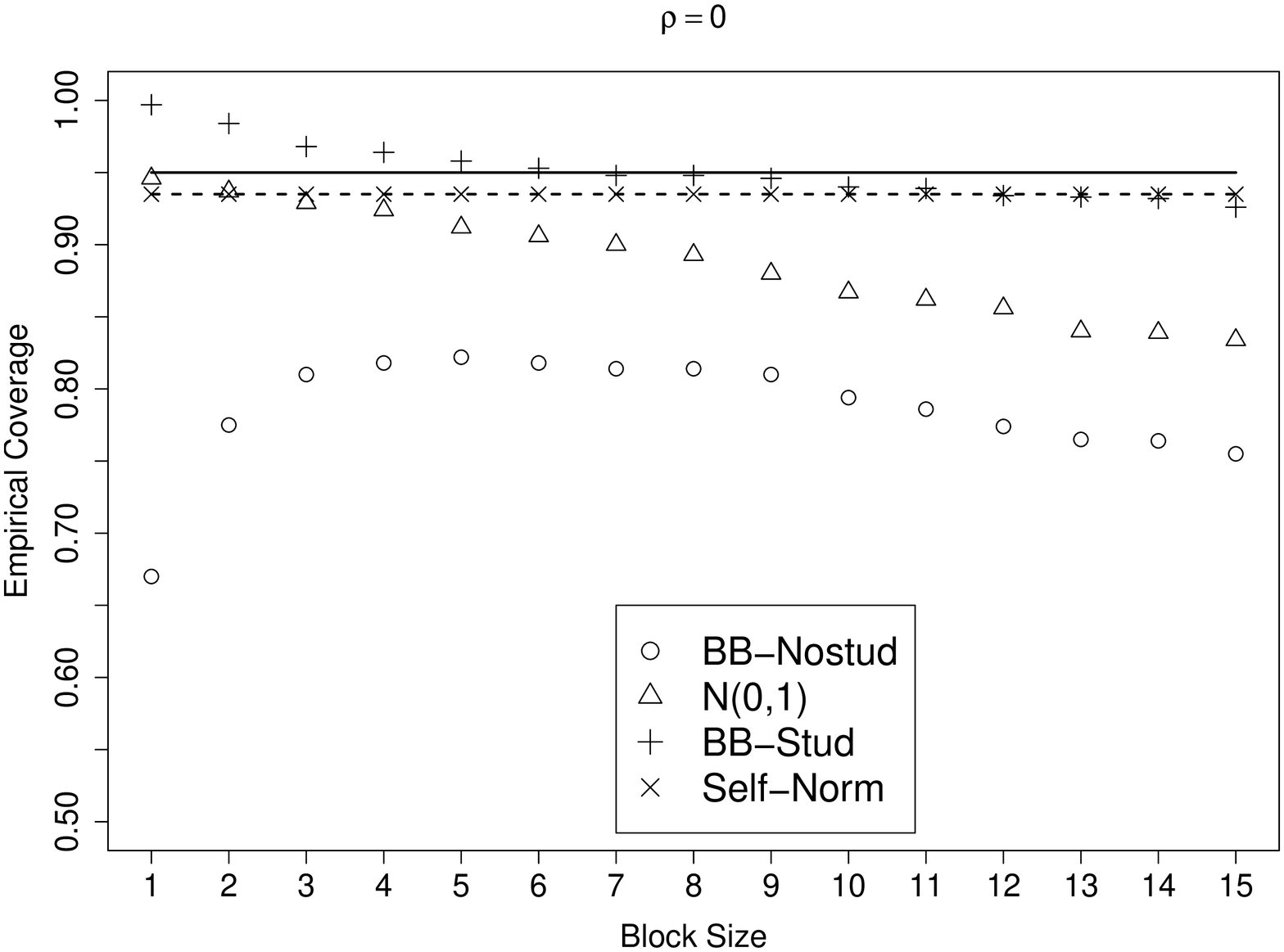}}
{\includegraphics[height=8cm,width=6cm,angle=270]{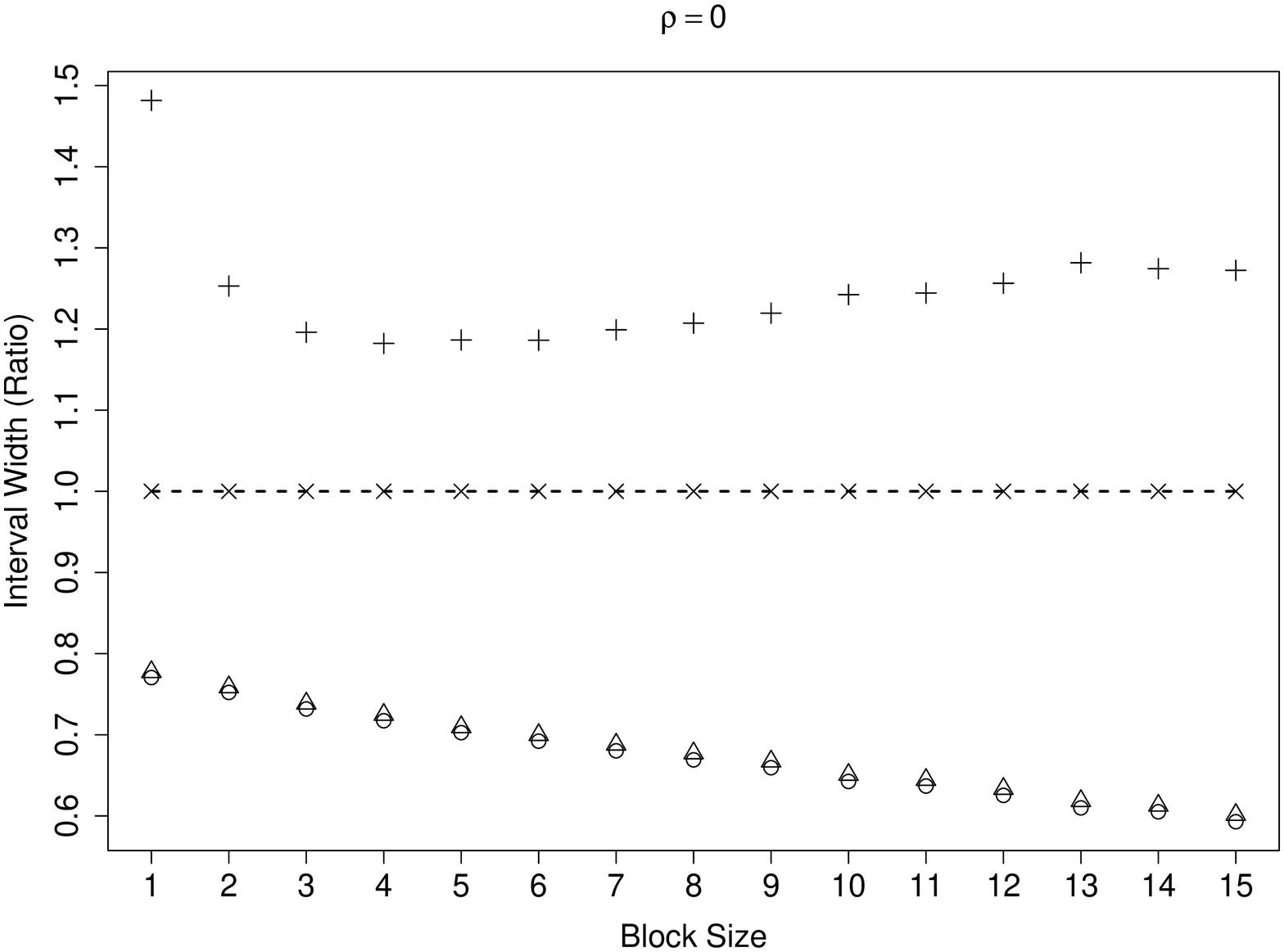}}
{\includegraphics[height=8cm,width=6cm,angle=270]{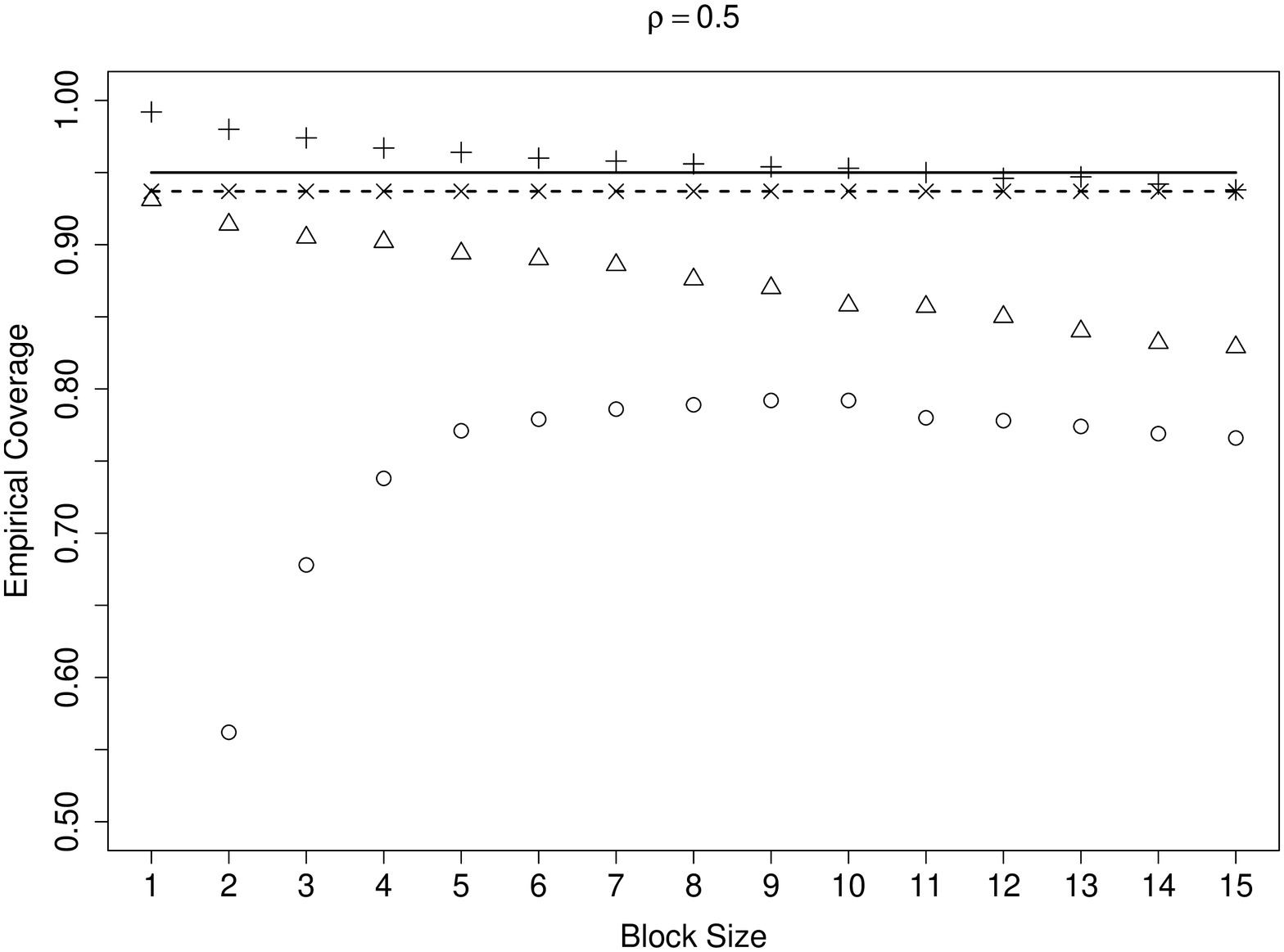}}
{\includegraphics[height=8cm,width=6cm,angle=270]{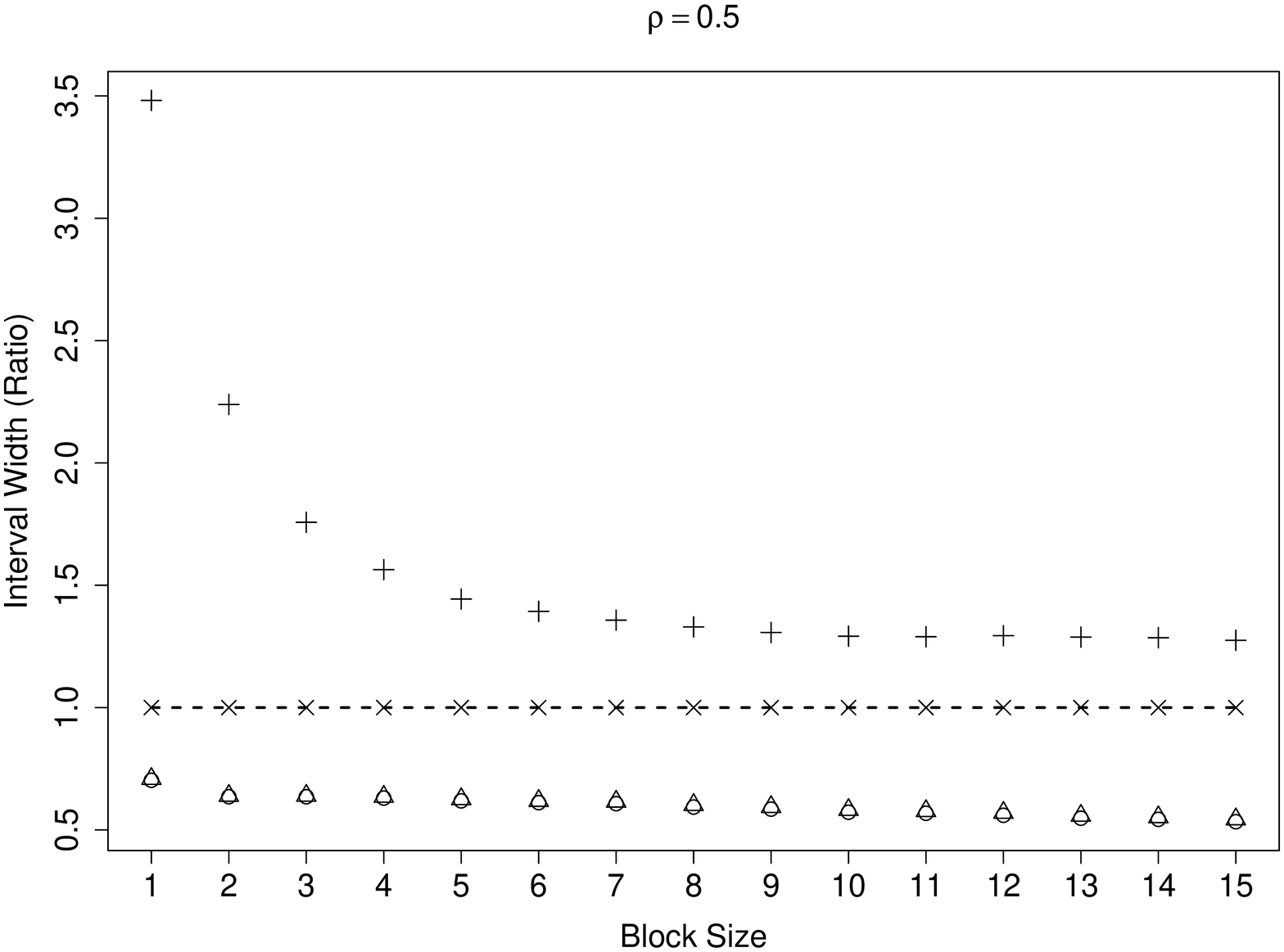}}
{\includegraphics[height=8cm,width=6cm,angle=270]{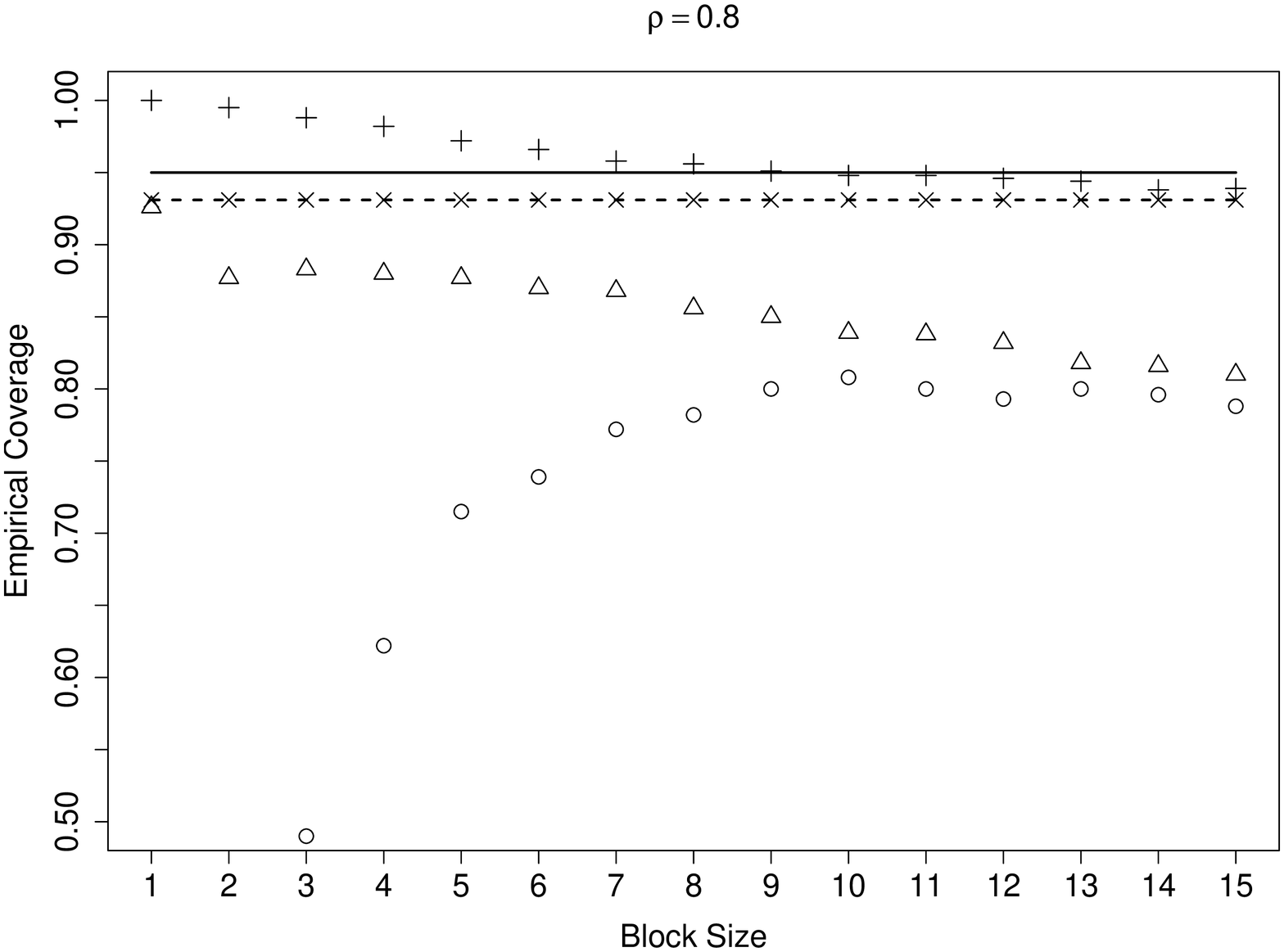}}
{\includegraphics[height=8cm,width=6cm,angle=270]{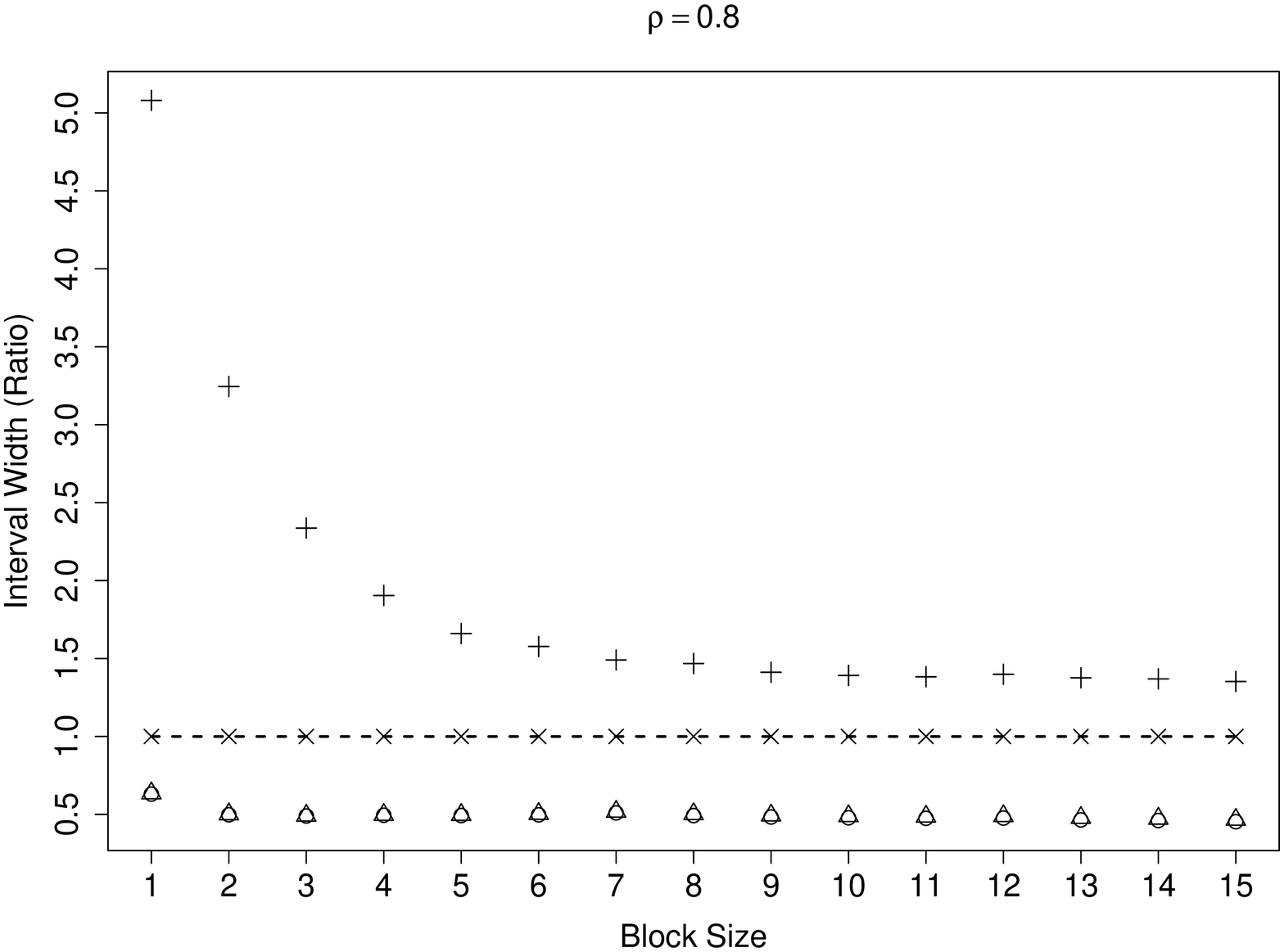}}
\label{fig:acf}
\end{center}

\end{figure}

\begin{figure}
\caption{Empirical coverage probabilities (left panel) and  ratios of the interval widths over that delivered by
 the self-normalized method (right panel) for $F(\pi/2)/F(\pi)$. Sample size $n=50$ and number of replications is 500. }
\begin{center}
{\includegraphics[height=8cm,width=6cm,angle=270]{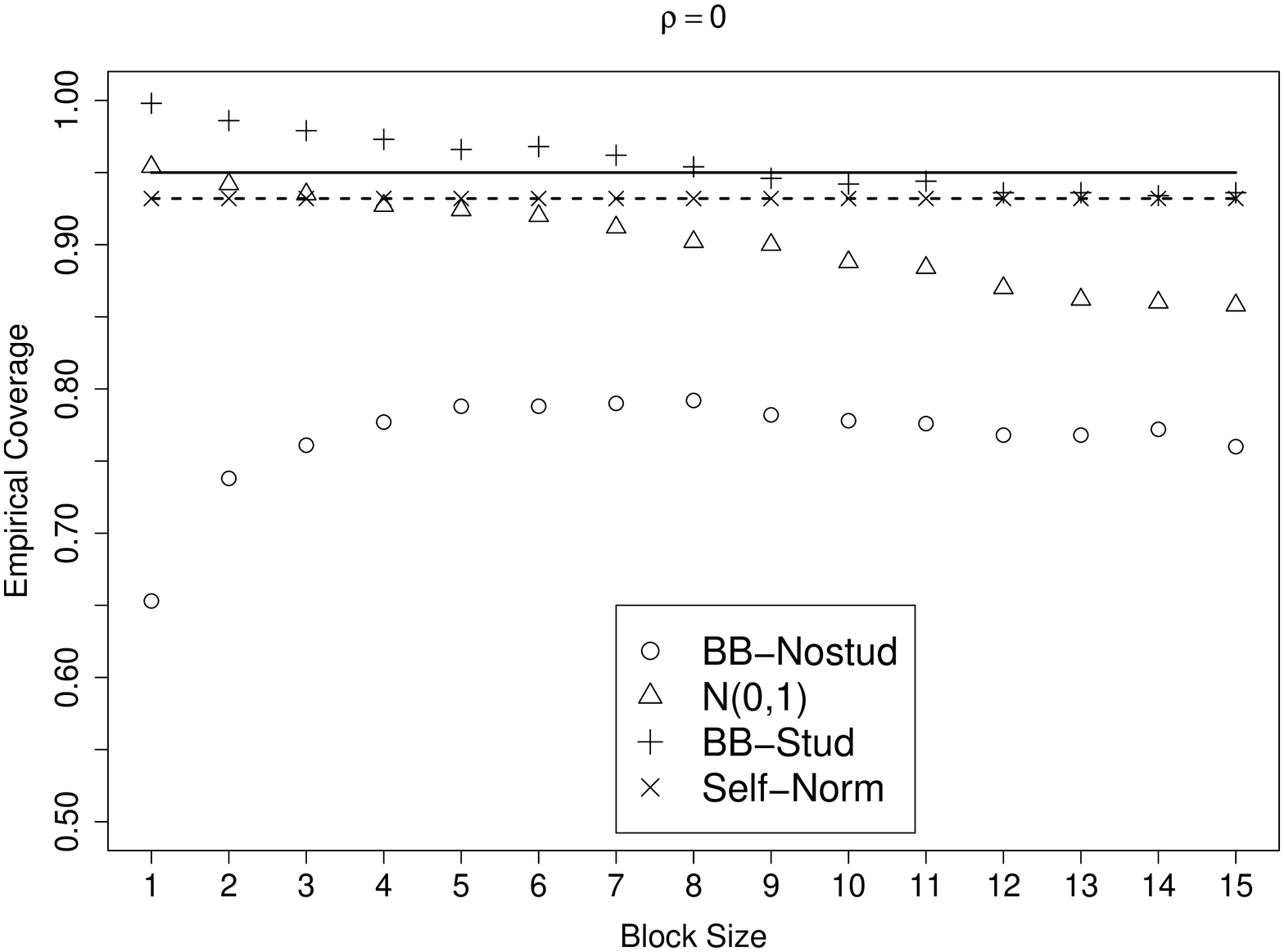}}
{\includegraphics[height=8cm,width=6cm,angle=270]{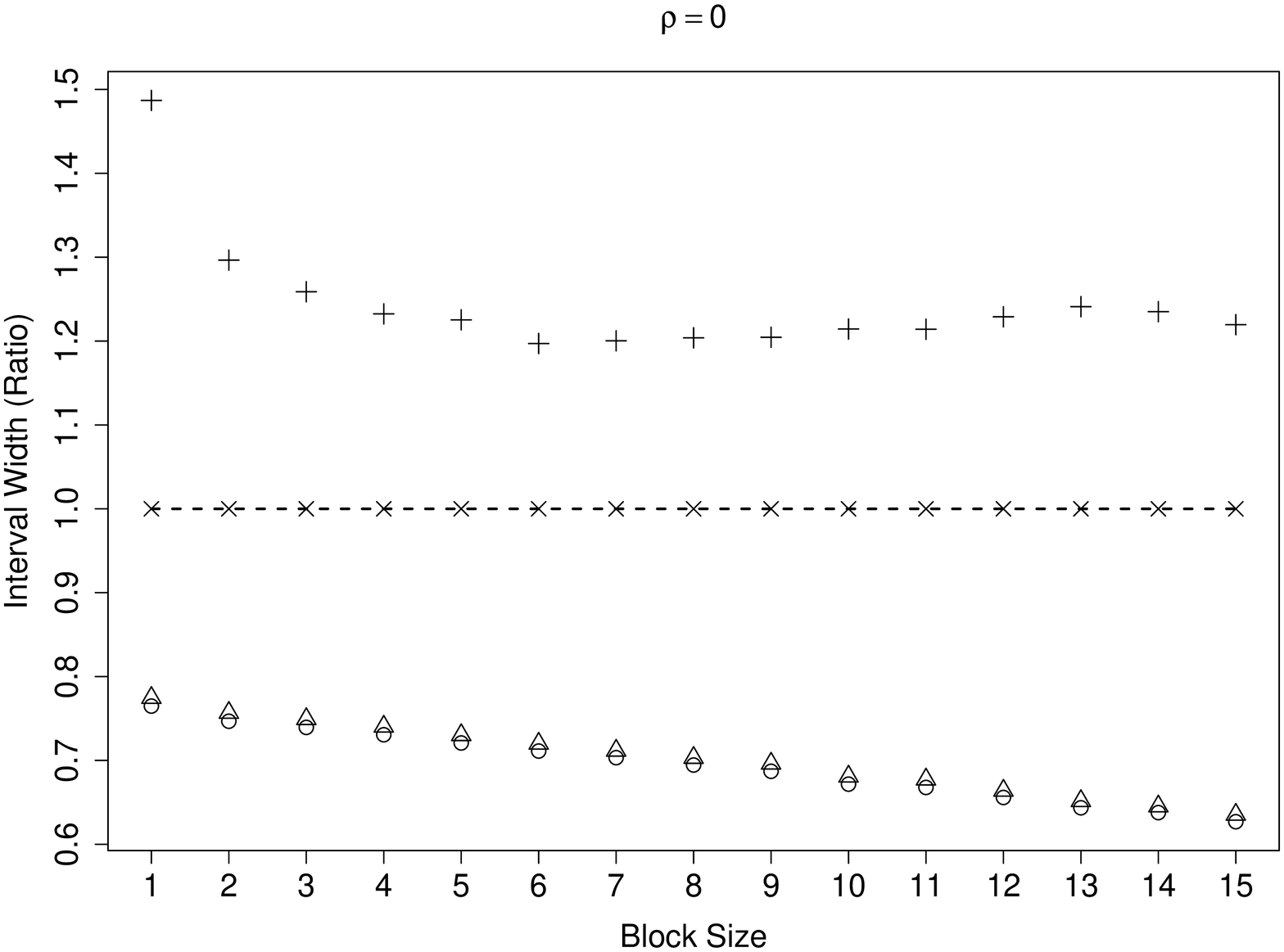}}
{\includegraphics[height=8cm,width=6cm,angle=270]{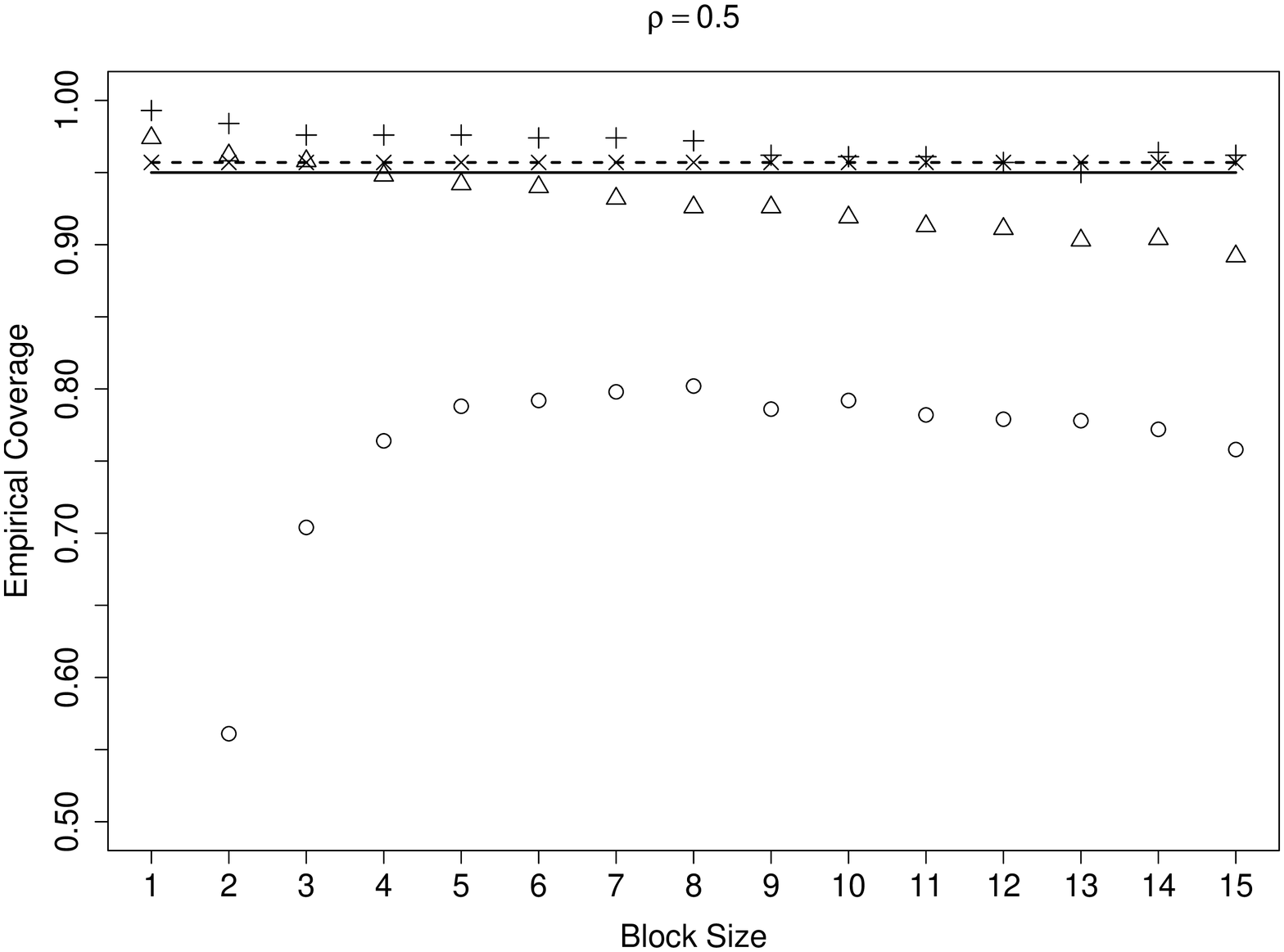}}
{\includegraphics[height=8cm,width=6cm,angle=270]{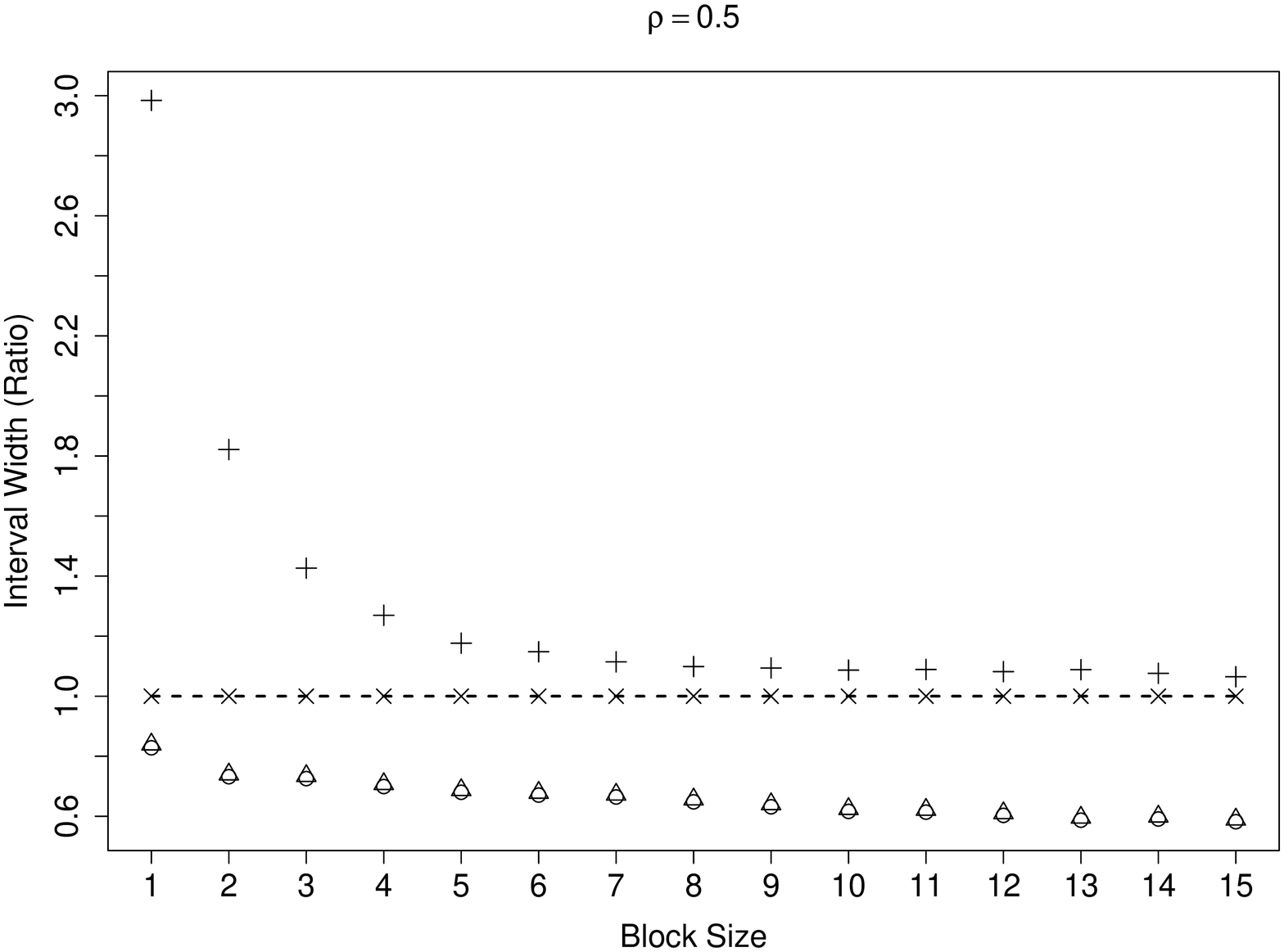}}
{\includegraphics[height=8cm,width=6cm,angle=270]{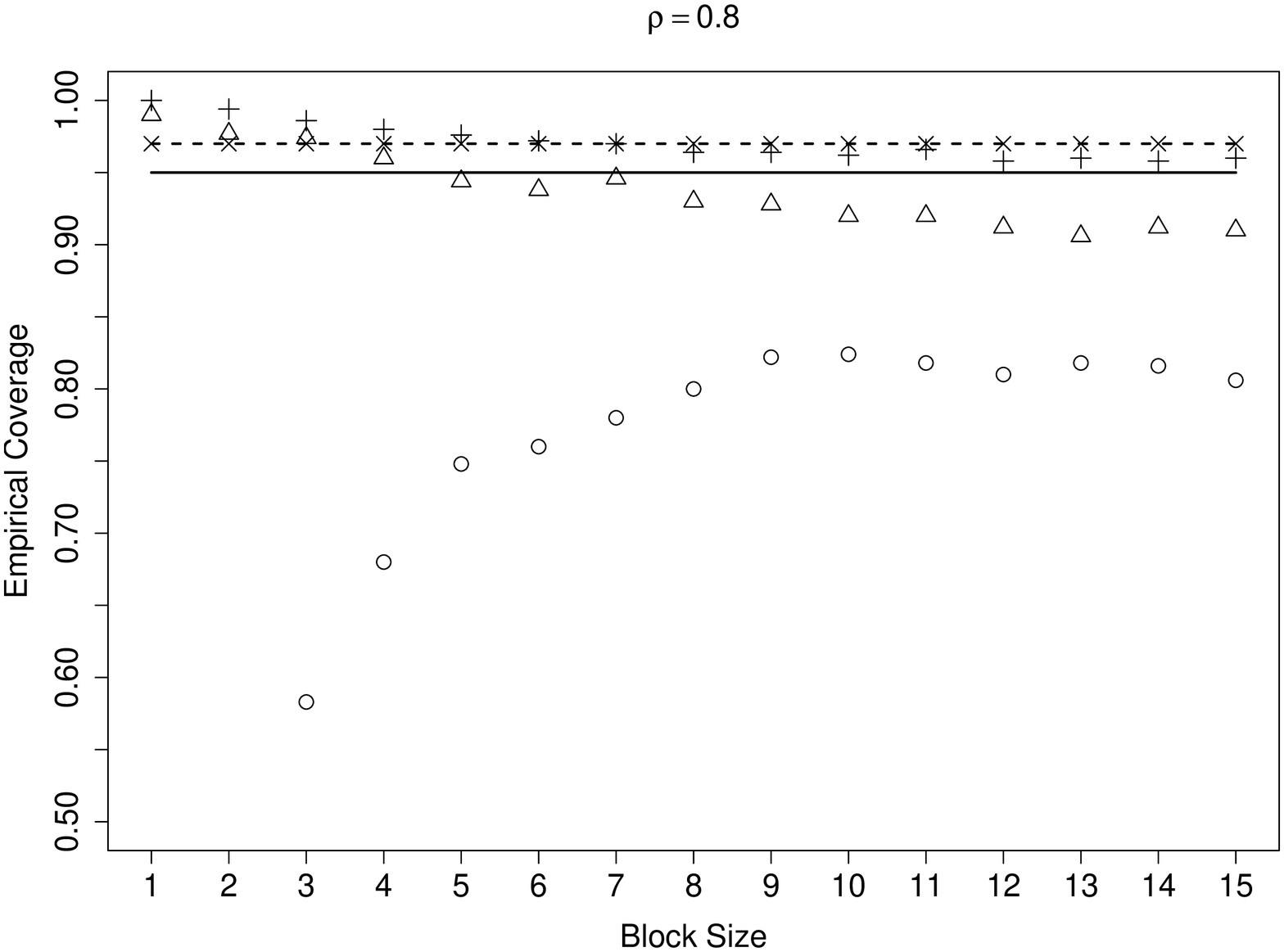}}
{\includegraphics[height=8cm,width=6cm,angle=270]{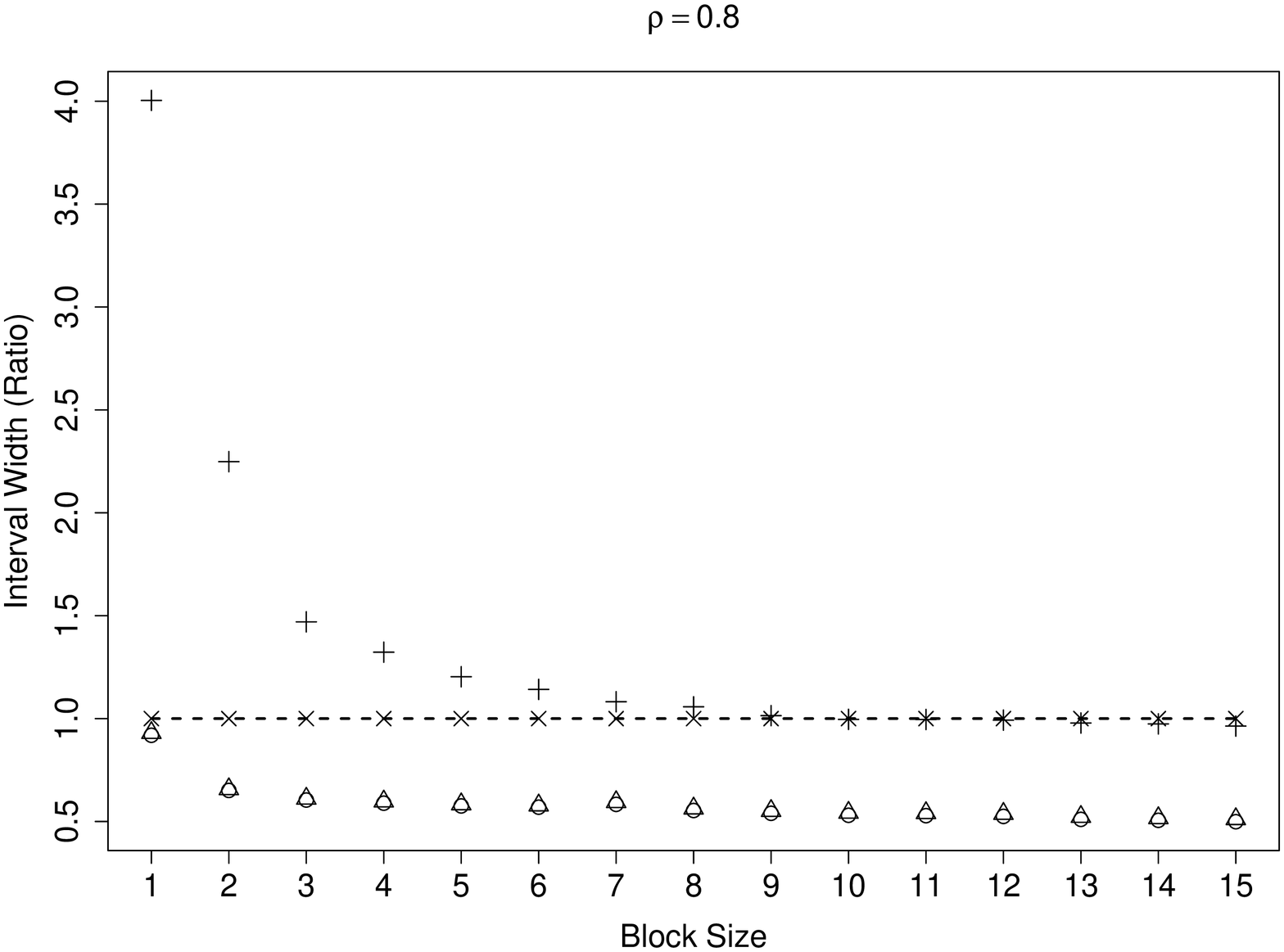}}
\label{fig:spec}
\end{center}

\end{figure}

In Figures~\ref{fig:mean}-\ref{fig:spec}, we plot the empirical coverage probabilities and the ratio of the mean interval widths
 over that delivered by the self-normalized method for $\E(X_1)$, $med(X_1)$, $\rho(1)$ and $F(\pi/2)/F(\pi)$ respectively.
The symbols ``BB-Nostud", ``$N(0,1)$", ``BB-Stud" and ``Self-Norm" in the figures correspond to the schemes  
(1)-(4) that were described above.
For both $\E(X_1)$ and $med(X_1)$, Figures~\ref{fig:mean} and~\ref{fig:med} show that all methods lead to undercoverage.
 The coverages for the moving block bootstrap without Studentizing are comparable with the normal approximation for $\E(X_1)$, but are noticeably
  inferior to the normal approximation uniformly in the block sizes that were examined for $med(X_1)$. The coverages for both methods deteriorate quickly as the correlation strengthens.
  For the moving block bootstrap with inefficient Studentizing, it shows very good coverages across the range of block sizes and outperforms the self-normalized method for all block sizes when $\rho=0.8$. As far as the length of the intervals is concerned,  the intervals corresponding to  the normal approximation and moving block bootstrap without Studentizing are of similar widths and are shorter than  the self-normalization-based intervals. This is consistent with the loss of the local asymptotic power for the self-normalized method (Lobato 2001), as a test statistic that corresponds to a wider interval  tends to be less sensitive to the local alternatives. Also note that  the intervals that are  delivered by the moving block bootstrap with inefficient Studentizing are wider than those by the self-normalized method uniformly in the block sizes.

 \centerline{Please insert Figures~\ref{fig:mean}\&\ref{fig:med} here!}

   From Figures~\ref{fig:acf}
and~\ref{fig:spec}, we see that the empirical coverages for $\rho(1)$ and $F(\pi/2)/F(\pi)$ that are delivered by the self-normalized method are fairly close to the nominal level. Although  the bootstrap with inefficient Studentizing produces apparent overcoverage and very wide intervals with small block sizes,  it is still possible to achieve a better coverage than the self-normalized method with suitably chosen block sizes. The normal approximation is again seen to be superior to the moving block bootstrap without Studentizing uniformly in the block sizes, although its coverages deviate from the nominal level when the block size increases. By contrast, the moving block bootstrap with inefficient Studentizing appears to be less sensitive to the choice of block size.

 \centerline{Please insert Figures~\ref{fig:acf}\&\ref{fig:spec} here!}

  For the inference of autocorrelations (e.g. $\rho(1)$), the moving block bootstrap method (without Studentizing), along with other nonparametric resampling methods, was advocated by Romano and Thombs (1996). Although it has been justified theoretically, the simulation results here suggest that the moving block bootstrap method   without Studentizing is not a good choice owing to its poor coverage.
The normal approximation, which involves the choice of block size in its variance estimation,  is also not recommended because of its sensitivity to the block size selection and unsatisfactory coverages. In comparison, the self-normalized method has reasonably good coverage and is free of any tuning parameters. In accordance with Goncalves and Vogelsang (2008), the block bootstrap (with inefficient Studentizing) can further improve the coverage of the self-normalized approach with suitable choice of block length. However, to achieve this slight improvement in coverage, we need to pay a computational cost and  the resulting interval tends to be wider.

\section{Conclusions}
\label{sec:con}

In this article, a new approach is proposed to constructing confidence intervals (regions) for quantities in time series.
The appealing features of the proposed self-normalized approach can be summarized as follows: (a) It is based on an asymptotically pivotal statistic  and does not
involve any user-chosen numbers; (b) It is easy to implement, since in general the calculation of recursive estimates requires only additional computation without the need to design any new algorithms; (c) It is broadly applicable to approximately linear statistics that are functionals of empirical distributions of fixed dimension and their asymptotically equivalent variants. Additionally, the theory can be extended to cover  spectral mean and its normalized version, which are  important quantities in time series. On the basis of the encouraging  finite sample performance that was  presented in Section~\ref{sec:sim} and the above nice characteristics,  we  recommend this procedure  to practitioners as a useful inference tool for routine use, such as obtaining the confidence interval of the lag 1 auto-correlation.

Simulation results suggest that our tuning-parameter-free approach may be further improved by applying the moving block bootstrap to approximate the sampling distribution of the self-normalized statistic. However, the improved coverage over the self-normalized approach is not guaranteed and it critically depends on the choice of block size.
It would be interesting to come up with a sound data-dependent block size selection rule. The early proposals by Hall et al. (1995) and Politis et al. (1999) on block length selection may still work for the current problem but need more investigation.
In practice, if the user knows how to choose the data-dependent block size properly and can   afford the computational cost that is associated with the block bootstrap method and the selection of block size, he or she is certainly encouraged to use the slightly wider bootstrap-based interval. In general, there are still grounds for recommending the self-normalized approach for its simplicity, convenience and reasonably good coverage.  An interesting theoretical topic is to show that the block bootstrap can improve the self-normalization-based approximation in terms of the ERP. This remains an open problem. In addition, the extension of this method to spatial settings is  worthwhile but seems  not straightforward for irregular spatial data. This is currently under investigation.

\section{Appendix}

Throughout the appendix, the positive constant $C$ is generic and it may vary from line to line.

\noindent Proof of Proposition~\ref{prop:1}:   Note that
      \begin{eqnarray*}
      r_n^2\E|T_{B_n}(\rho_{B_n}^{r_n})-\hat{\theta}_{r_n}|^2=r_n^2\var\left\{\sum_{k=B_n}^{r_n-1} \hat{\gamma}_{r_n}(k) g_k\right\}+r_n^2\left[\sum_{k=B_n}^{r_n-1} \E \{\hat{\gamma}_{r_n}(k)\} g_k\right]^2=I+II,
      \end{eqnarray*}
      where the latter term $II$ is less than or equal to $r_n^2 (\sum_{k=B_n}^{r_n-1}|\gamma(k)| |g_k|)^2\le C n^2(\sum_{k=B_n}^{\infty}|\gamma(k)|)^2=o(n)$. Let $W_{tk}=X_tX_{t-k}$. Regarding term $I$, we have
      \begin{eqnarray*}
    I  &\le &2\var\left(\sum_{k=B_n}^{r_n-1}\sum_{t=1}^{r_n} W_{tk} g_k\right)+2\var\left(\sum_{k=B_n}^{r_n-1}\sum_{t=1}^{k}W_{tk}g_k\right)=I_1+I_2,
    \end{eqnarray*}
    where, by the argument that was used in the proof of Theorem 1 in Shao (2009), term $I_1$ is less than or equal to $Cn\sum_{j=B_n}^{\infty}g_j^2=o(n)$.
    Note that we have applied the fact that $f(\cdot)$ and $f_4(\cdot,\cdot,\cdot)$ are both bounded under condition (\ref{eq:summability}). Next, we write 
    \begin{eqnarray*}
    I_2/2&=&\sum_{k,k'=B_n}^{r_n-1}\sum_{t=1}^{k}\sum_{t'=1}^{k'}\cov(W_{tk},W_{t'k'})g_k g_{k'}\\
    &=&\sum_{k,k'=B_n}^{r_n-1}\sum_{t=1}^{k}\sum_{t'=1}^{k'}g_k g_{k'}\{\gamma(t-t')\gamma(t'-k'-t+k)+\gamma(t'-k'-t)\times\\
      &&\gamma(t'-t+k)+\cum(X_0,X_{-k},X_{t'-t},X_{t'-k'-t})\}=I_{21}+I_{22}+I_{23}.
    \end{eqnarray*}
    Let $\Pi^2=[-\pi,\pi]^2$ and $H_k(\lambda)=\sum_{t=1}^{k}e^{it\lambda}$. For term $I_{21}$, we have that
    \begin{eqnarray*}
    I_{21}&=&\int_{\Pi^2}\sum_{k,k'=B_n}^{r_n}\sum_{t=1}^{k}\sum_{t'=1}^{k'}g_k g_{k'} e^{i(t'-t)\lambda_1}f(\lambda_1) e^{i(t'-t-k'+k)\lambda_2} f(\lambda_2)d\lambda_1 d\lambda_2\\
    &=&\int_{\Pi^2}\sum_{k,k'=B_n}^{r_n}g_k e^{ik\lambda_2} H_k(-\lambda_1-\lambda_2) g_{k'}e^{-ik'\lambda_2}H_{k'}(\lambda_1+\lambda_2)f(\lambda_1)f(\lambda_2)d\lambda_1 d\lambda_2.
    \end{eqnarray*}
    By the Cauchy-Schwarz inequality, we have that
    \begin{eqnarray*}
    |I_{21}|&\le&C\int_{\Pi^2}\left|\sum_{k=B_n}^{r_n}g_k e^{ik\lambda_2} H_k(w)\right|^2dw d\lambda_2.
    \end{eqnarray*}
    Summation by parts yields
     \[\sum_{k=B_n}^{r_n}g_k e^{ik\lambda_2} H_k(w)=H_{r_n}(w)\sum_{k=B_n}^{r_n}g_k e^{ik\lambda_2}-\sum_{k=B_n}^{r_n-1}e^{i(k+1)w}\sum_{h=B_n}^{k} g_h e^{ih\lambda_2},\]
    and consequently,
    \begin{eqnarray*}
    |I_{21}|&\le&C\int_{\Pi^2}|H_{r_n}(w)|^2\left|\sum_{k=B_n}^{r_n}g_k e^{ik\lambda_2}\right|^2d\lambda_2 dw
    +C\int_{\Pi^2}\left|\sum_{k=B_n}^{r_n-1}e^{i(k+1)w}\sum_{h=B_n}^{k} g_h e^{ih\lambda_2}\right|^2d\lambda_2 dw\\
    &\le&Cn\sum_{k=B_n}^{\infty}g_k^2+C\int_{\Pi^2}\sum_{k=B_n}^{r_n-1}\left|\sum_{h=B_n}^{k} g_h e^{ih\lambda_2}\right|^2d\lambda_2\le Cn \sum_{k=B_n}^{\infty}g_k^2=o(n).
    \end{eqnarray*}
     By the same argument, we can show that term $|I_{22}|=o(n)$. As for term $I_{23}$, we have
     \begin{eqnarray*}
     |I_{23}|&\le&C\sup_{k\ge B_n} g_k^2 \sum_{k,k'=B_n}^{r_n-1}\sum_{t=1}^{r_n}\sum_{t'=1}^{r_n}|\cum(X_0,X_{-k},X_{t'-t},X_{t'-k'-t})|\\
     &\le&C n\sup_{k\ge B_n} g_k^2 \sum_{k,k',h\in\Z}|\cum(X_0,X_{-k},X_h,X_{h-k'})|=o(n)
     \end{eqnarray*}
     under condition (\ref{eq:summability}). Therefore, term $|I|=o(n)$ and $r_n^2\E|T_{B_n}(\rho_{B_n}^{r_n})-\hat{\theta}_{r_n}|^2=o(n)$.
     Finally, we note that
\begin{eqnarray*}
     r_n^2 \E|R_{r_n n}|^2&=&\sum_{k,k'=0}^{B_n-1}\sum_{t=1}^{k}\sum_{t'=1}^{k'}g_{k'} g_k\cov(W_{tk},W_{t'k'})\\
      &=&\sum_{k,k'=0}^{B_n-1}\sum_{t=1}^{k}\sum_{t'=1}^{k'}g_{k'} g_k\{\gamma(t-t')\gamma(t'+k'-t-k)\\
      &&+\gamma(t'+k'-t)\gamma(t'-t-k)+\cum(X_0,X_{k},X_{t'-t},X_{t'+k'-t})\}.
      \end{eqnarray*}
      Applying the same argument as used in term $I_{21}$, we can derive that the first two terms in the preceding display are $O(B_n)=o(n)$, and the last term is
      bounded by
      \[C \sum_{k,k'=0}^{B_n-1}\sum_{t=1}^{k}\sum_{t'=1}^{k'}|\cum(X_0,X_{k},X_{t'-t},X_{t'+k'-t})|=O(B_n)=o(n).\]

It is easy to see that the above bound holds uniformly in $r\in (0,1]$. This completes the proof.

\qed

\bigskip

\centerline {\bf\large\sc References}

\par\noindent\hangindent2.3em\hangafter 1
{Billingsley, P.} (1968) {\it Convergence of Probability Measures}, Wiley.

\par\noindent\hangindent2.3em\hangafter 1
{Brillinger, D. R.} (1969) Asymptotic properties of spectral
estimates of second order. {\it Biometrika}, {\bf 56}, 375-390.

\par\noindent\hangindent2.3em\hangafter 1
B\"uhlmann, P. (2002) Bootstraps for time series. {\it Statistical Science}, {\bf 17}, 52-72.

\par\noindent\hangindent2.3em\hangafter 1
B\"uhlmann, P., and K\"unsch, H. R. (1995) The blockwise bootstrap for general parameters of a stationary time series.
{\it Scand. J. Statist.}, {\bf 22}, 35-54.

\par\noindent\hangindent2.3em\hangafter 1
Bunzel, H., Kiefer, N. M. and Vogelsang, T. J. (2001) Simple robust testing of hypotheses in nonlinear models.
{\it J. Am. Statist. Assoc.}, {\bf 96}, 1088-1096.

\par\noindent\hangindent2.3em\hangafter 1
Carrasco, M., and Chen, X. (2002) Mixing and moment properties of various GARCH and
stochastic volatility models.  {\it Econometric Theory}, {\bf 18}, 17–39.

\par\noindent\hangindent2.3em\hangafter 1
{Chiu, S. T.} (1988) Weighted least squares estimators on the
frequency domain for the parameters of a time series. {\it Ann.
 Statist.}, {\bf 16}, 1315-1326.

\par\noindent\hangindent2.3em\hangafter 1
{Dahlhaus, R.} (1985) Asymptotic normality of spectral estimates.
{\it J. Mult. Anal.}, {\bf 16}, 412-431.

\par\noindent\hangindent2.3em\hangafter 1
{Dahlhaus, R.}, and {Janas, D.} (1996) A frequency domain bootstrap
for ratio statistics in time series analysis. {\it  Ann. Statist.}, {\bf 24}, 1934-1963.

\par\noindent\hangindent2.3em\hangafter 1
{Davidson, J.} (2002) Establishing conditions for the functional central limit theorem in nonlinear and semiparametric time series processes.
 {\it J. Econometrics.}, {\bf 106}, 243-269.

\par\noindent\hangindent2.3em\hangafter 1
{Franke, J.}, and {H\"ardle, W.} (1992) On bootstrapping kernel
spectral estimates. {\it Ann.   Statist.}, {\bf 8}, 121-145.

\par\noindent\hangindent2.3em\hangafter 1
{Goncalves, S.}, and {Vogelsang, T. J.}  (2008) Block bootstrap HAC robust tests: the sophistication of the naive bootstrap.
 Working paper, University of Montreal and Michigan State University.

\par\noindent\hangindent2.3em\hangafter 1
Hall, P., Horowitz, J. L., and Jing, B.-Y. (1995) On blocking rules for the bootstrap with dependent data. {\it Biometrika},
 {\bf 82}, 561-574.

\par\noindent\hangindent2.3em\hangafter 1
{Hampel, F.}, {Ronchetti, E.}, {Rousseeuw, P.}, and {Stahel, W.} (1986), {\it Robust Statistics: The Approach Based on Influence Functions},  New York: John Wiley.

\par\noindent\hangindent2.3em\hangafter 1
Herrndorf, N. (1984) A functional central limit theorem for weakly dependent sequences of random variables. {\it Ann.  Probab.},
 {\bf 12}, 141-153.

\par\noindent\hangindent2.3em\hangafter 1
{Jansson, M.} (2004) The error rejection probability of simple autocorrelation
 robust tests. {\it Econometrica}, {\bf 72}, 937-946.

\par\noindent\hangindent2.3em\hangafter 1
{Keenan, D. M.} (1987) Limiting behavior of functionals of
higher-order sample cumulant spectra. {\it Ann. Statist.},
{\bf 15}, 134-151.

\par\noindent\hangindent2.3em\hangafter 1
{Kiefer, N. M.}, and {Vogelsang, T. J.} (2002a) Heteroskedasticity-autocorrelation robust
standard errors using the Bartlett kernel without truncation. {\it Econometrica}, {\bf 70}, 2093-2095.

\par\noindent\hangindent2.3em\hangafter 1
{Kiefer, N. M.}, and {Vogelsang, T. J.} (2002b) Heteroskedasticity-autocorrelation robust testing using bandwidth equal to sample size.
{\it Econometric Theory}, {\bf 18}, 1350-1366.

\par\noindent\hangindent2.3em\hangafter 1
{Kiefer, N. M.}, and {Vogelsang, T. J.} (2005) A new asymptotic theory for heteroskedasticity-autocorrelation robust tests.
 {\it Econometric Theory}, {\bf 21}, 1130-1164.

\par\noindent\hangindent2.3em\hangafter 1
{Kiefer, N. M.}, {Vogelsang, T. J.}, and {Bunzel, H.} (2000) Simple robust testing of regression hypotheses. {\it Econometrica},
 {\bf 68}, 695-714.

\par\noindent\hangindent2.3em\hangafter 1
{Kitamura, Y.} (1997) Empirical likelihood methods with weakly dependent processes. {\it  Ann. Statist.},
 {\bf 25}, 2084-2102.

\par\noindent\hangindent2.3em\hangafter 1
{Kreiss, J. P.}, and {Paparoditis, E.} (2003) Autoregressive-aided
periodogram bootstrap for time series. {\it  Ann. Statist.},
 {\bf 31}, 1923-1955.

\par\noindent\hangindent2.3em\hangafter 1
{Kuan, C. M.}, and {Lee, W. M.} (2006) Robust M tests without consistent estimation of the asymptotic covariance matrix.
{\it J. Am. Statist. Assoc.}, {\bf 101}, 1264-1275.

\par\noindent\hangindent2.3em\hangafter 1
{K\"unsch, H.} (1989) The jackknife and the bootstrap for general stationary observations. {\it  Ann. Statist.},
 {\bf 17}, 1217-1241.

\par\noindent\hangindent2.3em\hangafter 1
Lai, T.L., de la Pena, V. and Shao, Q. M. (2009) {\it  Self-normalized Processes: Theory and Statistical Applications}, Springer Series in Probability and its Applications,  Springer-Verlag , New York.

\par\noindent\hangindent2.3em\hangafter 1
{Lee, W. M.} (2007) Robust M tests using kernel-based estimators with bandwidth equal to sample size. {\it Economics Letters},
 {\bf 96}, 295-300.

\par\noindent\hangindent2.3em\hangafter 1
{Liu, R. Y.}, and {Singh, K.} (1992) Moving block jackknife and bootstrap capture weak dependence. In {\it Exploring
the Limits of Bootstrap}, Ed. R. LePage and L. Billard, pp. 225-248. New York: John Wiley.

\par\noindent\hangindent2.3em\hangafter 1
{Lobato, I. N.} (2001) Testing that a dependent process is
uncorrelated. {\it J. Am. Statist.
Assoc.}, {\bf 96}, 1066-1076.

\par\noindent\hangindent2.3em\hangafter 1
{Lobato, I. N.}, {Nankervis, J. C.} and {Savin, N. E.}  (2002)
Testing for zero autocorrelation in the presence of statistical
dependence. {\it Econometric Theory}, {\bf 18}, 730-743.

\par\noindent\hangindent2.3em\hangafter 1
{McElroy, T.} and {Politis, D. N.} (2007) Computer-intensive rate estimation, diverging statistics, and scanning.
{\it Ann. Statist.}, {\bf 35}, 1827-1848.

\par\noindent\hangindent2.3em\hangafter 1
Newey, W. K., and West, K. D. (1994) Automatic lag selection in covariance matrix estimation.
{\it Review of Economic Studies}, {\bf 61}, 631-653.

\par\noindent\hangindent2.3em\hangafter 1
{Nordman, D. J.}, and {Lahiri, S. N.} (2006) A frequency domain
empirical likelihood for short- and long-range dependence. {\it
Ann. Statist.}, {\bf 34}, 3019-3050.

\par\noindent\hangindent2.3em\hangafter 1
Pham, T. D., and Tran, L. T. (1985) Some mixing properties of time series models.
{\it Stochast. Process. Appl.}, {\bf 19}, 297-303.

\par\noindent\hangindent2.3em\hangafter 1
Pham, T. D. (1986) The mixing property of bilinear and generalised random coefficient
autoregressive models. {\it Stochast. Process. Appl.}, {\bf 23}, 291-300.

\par\noindent\hangindent2.3em\hangafter 1
Phillips, P. C. B. (1987) Time series regression with a unit root. {\it Econometrica}, {\bf 55}, 277-301.

\par\noindent\hangindent2.3em\hangafter 1
Phillips, P. C. B., Sun, Y. and Jin, S. (2004) Improved HAR inference using power kernels without truncation.
Working paper, Department of Economics, Yale University.

\par\noindent\hangindent2.3em\hangafter 1
Phillips, P. C. B., Sun, Y. and Jin, S. (2006) Spectral density estimation and robust hypothesis testing using
steep origin kernels without truncation. {\it Int. Econ. Rev.}, {\bf 47}, 837-894.

\par\noindent\hangindent2.3em\hangafter 1
Phillips, P. C. B., Sun, Y. and Jin, S. (2007) Long run variance estimation and robust regression testing using sharp origin kernels with no truncation. {\it J. Stat. Plan. Infer.},  {\bf 137},  985-1023.

\par\noindent\hangindent2.3em\hangafter 1
{Politis, D. N.}, {Romano, J. P.}, and {Wolf, M.} (1999), {\it
Subsampling}, Springer-Verlag, New York.

\par\noindent\hangindent2.3em\hangafter 1
{Romano, J. L.}, and {Thombs, L. A.} (1996) Inference for
autocorrelations under weak assumptions. {\it J. Am. Statist. Assoc.}, {\bf 91}, 590-600.

\par\noindent\hangindent2.3em\hangafter 1
{Rosenblatt, M.} (1985), {\it Stationary Sequences and Random
Fields}, Birkh\"auser, Boston.

\par\noindent\hangindent2.3em\hangafter 1
Shao, X. (2009) Confidence intervals for spectral mean and ratio statistics. {\it Biometrika}, {\bf 96}, 107-117.

\par\noindent\hangindent2.3em\hangafter 1
Sun, Y., Phillips, P. C. B. and Jin, S. (2008) Optimal bandwidth selection in heteroscedasticity-autocorrelation robust testing. {\it Econometrica}, {\bf 76}, 175-194.

\par\noindent\hangindent2.3em\hangafter 1
{Taniguchi, M.} (1982) On estimation of the integrals of the 4th
order cumulant spectral density. {\it Biometrika}, {\bf 69}, 117-122.

\par\noindent\hangindent2.3em\hangafter 1
 Velasco, C., and Robinson, P. M. (2001) Edgeworth expansions for spectral density estimates and Studentized sample mean.
 {\it Econometric Theory}, {\bf 17}, 497-539.

\newpage

\begin{table}
 \caption{ Empirical rejection percentages  for the eight models at $5\%$ and $10\%$ levels when (a) $n=100$ and (b) $n=500$. The rows (i), (ii) and (iii) correspond to the results for $T_K$, $\tilde{T}_K$ and $\tilde{Q}_K$ respectively.
The number of replications is 5000. The largest standard error is $0.53\%$.}

\begin{scriptsize}
\begin{center}
\begin{tabular}{cc|c|cccccccc}
\hline \hline
(1a)&&&$N(0,1)$&Student&LogNormal&RT&Hetero&No-MDS&GARCH&Bilinear\\
\hline
$K=1$&5$\%$&(i)&4.7&4.6&4.0&3.8&3.2&2.8&4.6&5.1\\
&&(ii)&4.8&5.1&4.3&3.8&3.3&3.2&4.6&5.5\\
&&(iii)&4.0&3.7&1.9&2.4&1.3&2.4&4.1&6.8\\
&10$\%$&(i)&9.8&10.2&9.0&8.6&7.8&7.4&10.2&10.9\\
&&(ii)&9.9&10.5&9.5&9.5&8.2&8.0&10.2&11.6\\
&&(iii)&9.1&9.5&6.4&7.1&5.3&7.2&9.2&13.6\\
$K=3$&5$\%$&(i)&5.1&3.9&1.7&2.0&1.1&0.9&5.2&4.4\\
&&(ii)&7.0&5.6&3.3&3.4&2.4&1.7&6.7&7.1\\
&&(iii)&3.5&2.6&1.3&1.0&0.1&1.8&3.4&3.7\\
&10$\%$&(i)&10.0&9.3&5.0&6.0&4.2&2.7&10.9&10.1\\
&&(ii)&13.8&10.9&7.7&8.1&6.2&4.9&13.3&13.7\\
&&(iii)&8.7&7.2&4.1&2.2&0.5&2.9&8.2&8.8\\
$K=5$&$5\%$&(i)&4.2&2.9&0.9&0.8&0.5&0.3&3.4&3.2\\
&&(ii)&8.9&6.4&3.7&2.7&1.7&1.3&8.1&7.3\\
&&(iii)&1.9&1.5&1.6&1.0&2.4&2.9&1.8&2.0\\
&10$\%$&(i)&9.1&6.4&2.8&2.5&1.7&1.1&8.0&7.2\\
&&(ii)&16.5&12.9&7.9&6.8&4.6&3.4&14.6&14.9\\
&&(iii)&5.4&4.2&3.1&1.9&3.2&3.6&5.0&5.5\\
\hline
\end{tabular}

\begin{tabular}{cc|c|cccccccc}
\hline
(1b)&&&$N(0,1)$&Student&LogNormal&RT&Hetero&No-MDS&GARCH&Bilinear\\
\hline
$K=1$&5$\%$&(i)&5.6&4.7&5.4&4.6&4.0&4.4&5.3&4.9\\
&&(ii)&5.7&4.6&5.5&4.7&3.9&4.4&5.2&5.0\\
&&(iii)&5.5&4.6&5.5&4.4&3.5&4.0&4.8&6.7\\
&10$\%$&(i)&10.9&10.1&10.8&10.2&9.0&9.7&10.3&9.8\\
&&(ii)&10.9&10.5&10.6&10.2&9.3&10.0&10.5&10.0\\
&&(iii)&10.8&9.9&11.4&9.8&8.7&9.8&10.2&12.3\\
$K=3$&5$\%$&(i)&5.0&4.9&4.7&4.4&3.2&3.6&5.3&5.9\\
&&(ii)&5.3&5.3&5.0&4.9&3.5&4.1&5.7&6.2\\
&&(iii)&4.7&4.1&3.2&2.5&0.4&1.2&4.6&5.3\\
&10$\%$&(i)&10.1&10.2&9.4&9.9&8.2&9.1&10.3&11.0\\
&&(ii)&11.2&10.4&10.3&10.6&8.4&9.6&11.2&11.9\\
&&(iii)&9.9&8.8&7.5&6.1&1.7&3.8&10.0&9.5\\
$K=5$&$5\%$&(i)&5.6&4.9&3.4&3.8&3.2&2.5&5.7&5.5\\
&&(ii)&6.8&5.8&4.4&4.7&3.6&3.3&6.3&6.5\\
&&(iii)&4.6&3.0&2.1&1.4&0.2&0.7&4.8&4.3\\
&10$\%$&(i)&11.1&9.8&8.0&8.4&6.8&6.2&11.1&11.0\\
&&(ii)&12.8&11.3&9.5&9.8&7.6&7.7&12.4&12.4\\
&&(iii)&9.5&7.7&5.4&4.3&1.2&1.6&9.5&8.5\\
\hline\hline
\end{tabular}
\label{tb:table1}
\end{center}

\end{scriptsize}
\end{table}

\newpage

\begin{table}
 \caption{(a) Size-adjusted power (in percent) under the alternative  AR$(1)$-GARCH$(1,1)$ (shown in (a)) and
 the  AR$(1)$ with innovations
 following the  bilinear model (8) (shown in (b))  at  $5\%$  and $10\%$ levels. The rows (i), (ii) and (iii) correspond to the results for $T_K$, $\tilde{T}_K$ and $\tilde{Q}_K$ respectively. The number of replications is 5000. The largest standard error is $0.71\%$.}

\begin{footnotesize}
\begin{center}
\begin{tabular}{cc|c|ccccc|ccccc}
\hline \hline
&&&&&(2a)&&&&&(2b)&&\\
&$\rho$&&0.1&0.2&0.3&0.4&0.5&0.1&0.2&0.3&0.4&0.5\\
\hline
$K=1$&5$\%$&(i)&11.4&29.7&52.4&69.0&78.2&7.8&19.4&35.9&51.7&61.9
\\
&&(ii)&11.1&29.3&50.9&67.0&75.7&8.1&19.0&35.9&51.3&61.6
\\
&&(iii)&11.8&38.1&67.6&85.7&90.0&7.3&20.2&39.9&56.1&62.0
\\
&10$\%$&(i)&18.5&42.8&67.2&82.3&89.2&14.6&30.7&51.7&68.5&77.7
\\
&&(ii)&18.4&41.9&66.6&81.4&88.3&14.1&30.2&51.7&67.8&76.6
\\
&&(iii)&21.3&53.5&81.7&94.8&97.3&13.7&32.6&55.8&71.8&77.7
\\
$K=3$&5$\%$&(i)&7.2&16.0&29.5&44.4&55.4&6.5&12.3&22.8&33.9&44.6
\\
&&(ii)&6.9&16.0&30.2&43.9&54.6&6.2&12.4&22.2&34.2&43.5
\\
&&(iii)&7.8&18.7&32.8&41.9&44.3&6.1&12.6&23.7&34.1&39.6
\\
&10$\%$&(i)&13.0&24.3&42.7&58.7&69.3&11.7&20.2&34.1&48.3&58.5
\\
&&(ii)&13.1&25.8&43.0&58.3&68.3&12.0&20.9&34.2&47.5&58.4
\\
&&(iii)&15.0&30.0&48.2&59.0&62.2&12.0&22.5&37.8&49.7&55.7
\\
$K=5$&$5\%$&(i)&7.0&13.4&24.8&37.9&47.7&5.4&9.6&16.9&25.3&33.4
\\
&&(ii)&6.2&12.8&23.6&36.3&46.1&5.9&10.8&18.2&27.3&35.5
\\
&&(iii)&8.4&16.0&27.3&36.9&43.8&6.3&11.0&19.7&28.6&35.1
\\
&10$\%$&(i)&11.6&21.5&36.1&49.9&60.6&11.3&17.5&27.7&39.0&48.7
\\
&&(ii)&12.6&21.5&36.6&50.1&60.6&12.2&19.0&28.5&40.9&50.0
\\
&&(iii)&14.0&25.0&40.2&51.8&58.9&11.9&19.9&31.7&42.6&50.9
\\
\hline\hline
\end{tabular}
\label{tb:table2}
\end{center}

\end{footnotesize}

\end{table}

\newpage

\begin{table}
 \caption{ Percentages of coverage  for the confidence
 interval for $\gamma(1)$ (shown in (a)) and $F(\pi/2)$ (shown in (b)) under models $M_1$-$M_6$.
 The number on the left-hand side of each column
  is the coverage percentage obtained from the method in Shao (2009),  the number in the square
 brackets stands for the coverage percentage delivered by the self-normalized method proposed in this paper, and the number in the curly brackets represents the coverage percentage for the efficiently Studentized method. The number of replications is 1000. The largest standard error is $1.42\%$.
}

\begin{footnotesize}

\begin{center}
\begin{tabular}{ccc|cc}
\hline \hline
(3a) &$n=150$&&$n=600$\\
$100(1-\alpha)\%$&$90\%$&$95\%$&$90\%$&$95\%$\\
$M_1$&81.4 [81.9] \{69.3\}& 87.3 [88.4] \{75.7\} &86.7 [86.9] \{73.3\}&92.1 [92.0] \{80.6\}\\
$M_2$&77.7 [77.7] \{68.7\} &83.4 [84.5] \{74.4\}&87.3 [86.8] \{78.9\}&92.3 [92.4] \{84.8\}\\
$M_3$&72.0 [71.9] \{63.0\}& 79.9 [78.3] \{68.4\}&81.9 [81.6] \{72.6\}&87.5 [87.3] \{79.0\}\\
$M_4$&85.2 [86.8] \{78.4\}&92.3 [91.8] \{85.5\}&87.3 [87.4] \{82.3\}&93.1 [93.2] \{88.5\}\\
$M_5$&83.3 [82.1] \{79.5\}& 88.4 [88.6] \{84.9\}&87.0 [86.4] \{85.0\}&93.8 [91.3] \{90.0\}\\
$M_6$&76.2 [75.4] \{69.4\}& 83.1 [82.1] \{76.2\}&82.5 [83.1] \{78.1\}&89.8 [88.9] \{84.1\}\\
\hline\hline
(3b)&$n=150$&&$n=600$\\
$100(1-\alpha)\%$&$90\%$&$95\%$&$90\%$&$95\%$\\
$M_1$& 82.3 [83.4]&88.0 [89.4] &86.2 [86.3]&92.3 [92.5]\\
$M_2$& 78.2 [80.6]& 84.5 [86.0]&87.1 [86.8]&91.6 [91.8]\\
$M_3$&74.7 [73.4]&80.1 [79.9]&83.2 [82.5]&89.4 [87.7]\\
$M_4$& 87.7 [88.1]&92.1 [93.0]&89.0 [89.1]&93.0 [93.7]\\
$M_5$&83.4 [82.3]& 88.7 [88.9]&85.9 [86.2]&92.0 [91.8] \\
$M_6$&78.6 [78.8]&84.2 [84.8]&83.4 [84.2]&89.4 [89.9]\\
\hline\hline
\end{tabular}
\label{tb:table3}
\end{center}
\end{footnotesize}

\end{table}

\newpage

\begin{table}
 \caption{Coverages  for the confidence interval of
 $\rho(1)$ (shown in (a)) and $F(\pi/2)/F(\pi)$ (shown in (b)) under  models $M_1$-$M_6$.
 The number on the left-hand side of each column
  is the coverage percentage obtained from the  method in Shao (2009);    the number
   in parentheses is the percentage that  produces  an empty
   interval by Shao's (2009) method; the number  in the square brackets stands for the coverage percentage
    delivered by the self-normalized method that is proposed in this paper; the number in braces represents the coverage percentage for the efficiently Studentized method. The number of replications is 1000. The largest standard error is $1.13\%$.
 }

\begin{scriptsize}
\begin{center}
\begin{tabular}{ccc|cc}
\hline \hline
(4a)&$n=150$&&$n=600$\\
$100(1-\alpha)\%$&$90\%$&$95\%$&$90\%$&$95\%$\\
$M_1$&85.1 (0.0) [90.4] \{89.5\}&92.2 (0.0) [95.6] \{93.8\}&88.8 (0.0) [89.7] \{93.6\}&94.7 (0.0) [94.9] \{96.0\}\\
$M_2$&85.7 (0.6) [91.9] \{93.6\}&90.6 (1.1) [95.4] \{95.8\}&91.4 (0.0) [91.5] \{96.8\}&95.3 (0.0) [96.4] \{98.5\}\\
$M_3$&86.6 (0.5) [87.5] \{88.3\}&89.8 (2.6) [93.2] \{92.2\}&89.1 (0.0) [88.1] \{93.7\}&93.6 (0.0) [93.2] \{96.2\}\\
$M_4$&89.0 (0.0) [87.9] \{90.8\}&94.1 (0.0) [95.1] \{94.3\}&89.1 (0.0) [89.3] \{94.5\}&94.0 (0.0) [94.4] \{96.7\}\\
$M_5$&90.6 (0.3) [88.7] \{96.2\}&94.1 (1.5) [93.7] \{97.8\}&88.4 (0.0) [89.8] \{99.1\}&93.6 (0.1) [94.7] \{99.7\}\\
$M_6$&89.8 (0.6) [86.1] \{88.1\}&93.0 (2.0) [90.2] \{92.1\}&89.8 (0.0) [87.6] \{91.4\}&95.3 (0.1) [93.3] \{94.9\}\\
\hline\hline
(4b)&$n=150$&&$n=600$\\
$100(1-\alpha)\%$&$90\%$&$95\%$&$90\%$&$95\%$\\
$M_1$&89.0 (0.0) [93.1]&93.8 (0.0) [97.0]&86.7 (0.0) [91.2]&93.1 (0.0) [95.3]\\
$M_2$&89.0 (0.2) [91.7]&93.2 (1.2) [95.8]&89.7 (0.0) [91.3]&94.6 (0.0) [96.3]\\
$M_3$&89.0 (1.2) [90.2]&93.1 (2.1) [94.7]&89.8 (0.0) [87.6]&94.9 (0.1) [93.1]\\
$M_4$&88.7 (0.0) [91.4]&94.8 (0.0) [95.3]& 89.5 (0.0) [89.3]& 94.3 (0.0) [95.2]\\
$M_5$&90.7 (0.1) [90.6]&94.8 (0.9) [95.8]&89.1 (0.0) [89.9]& 94.7 (0.1) [94.1]\\
$M_6$&91.1 (0.6) [88.7]&93.0 (2.3) [93.2]&90.8 (0.0) [88.4]& 95.7 (0.1) [93.3]\\
\hline\hline
\end{tabular}
\label{tb:table4}
\end{center}
\end{scriptsize}

\end{table}

\newpage

\begin{table}
 \caption{(a) Coverages  for the confidence interval for the median of $X_1$  out of 10000 replications
under  models $M_1$-$M_6$. The largest standard error is $0.34\%$. (b) Coverages for the confidence interval of autoregressive coefficients  based on the LAD regression for models $M_1$-$M_3$ and $M_7$-$M_9$. The number of replications is 1000. The largest standard error is $1.00\%$. }

\begin{footnotesize}

\begin{center}

\begin{tabular}{ccccc|cccccccc}
\hline \hline
&&(5a)&&&&&(5b)&\\
 &$n=150$&&$n=600$&&&$n=150$&&$n=600$ \\
$100(1-\alpha)\%$&$90\%$&$95\%$&$90\%$&$95\%$&&$90\%$&$95\%$&$90\%$&$95\%$\\
$M_1$&87.0&92.4 &89.1&94.2&$M_1$&92.5&95.4 &91.2&94.7\\
$M_2$&87.8&92.7&89.6&94.3&$M_2$&91.8&95.5&91.8&95.6\\
$M_3$&87.7&92.8 &88.4&93.5&$M_3$&92.7&96.5 &88.7&94.0\\
$M_4$&88.2&93.4&89.7&94.2&$M_7$&93.9&96.6&93.1&96.3\\
$M_5$&88.8&93.5 &89.2&93.9&$M_8$&94.8&97.7 &94.4&97.1\\
$M_6$&88.5&93.5 &89.6&94.3&$M_9$&93.1&95.9 &92.3&96.2\\
\hline\hline
\end{tabular}
\label{tb:table5}
\end{center}
\end{footnotesize}
\end{table}

\newpage

\newpage

\newpage

\newpage

\end{document}